\documentclass[12pt]{article}

\usepackage{newtxtext,newtxmath}

\usepackage{graphicx}

\usepackage[letterpaper,margin=1in]{geometry}

\renewenvironment{abstract}
	{\quotation}
	{\endquotation}

\date{}

\makeatletter
\renewcommand{\fnum@figure}{\textbf{Figure \thefigure}}
\renewcommand{\fnum@table}{\textbf{Table \thetable}}
\makeatother

\usepackage{scicite}

\usepackage{url}

\def\scititle{Analytical review of nanoplastic bioaccumulation data and a unified toxicokinetic model: from teleosts to human brain}
\title{\bfseries \boldmath \scititle}

\author{
	Alfonso M. Gañán-Calvo$^{1,2\ast}$,\\
	\small$^{1}$Dept. Ing. Aerospacial y Mecánica de Fluidos, Universidad de Sevilla, 41092 Sevilla, Spain.\and
	\small$^{2}$ENGREEN, Laboratory of Engineering for Energy and Environmental Sustainability,\\
     Universidad de Sevilla, 41092 Sevilla, Spain.\and
	\small$^\ast$Corresponding author. Email: amgc@us.es
	}

\begin{document}

\maketitle

\begin{abstract} \bfseries \boldmath
Nanoplastics (NPs) are increasingly detected in human blood and organs at concentrations reaching hundreds to thousands of parts per million, yet no quantitative framework has linked short-term experimental uptake kinetics to long-term, organ-specific accumulation. Here we analytically review the most reliable uptake and depuration datasets available in teleost fish using a sequential two-compartment toxicokinetic model that distinguishes systemic circulation from tissue-level retention. While anomalous, non-Markovian transport is expected at microscopic scales, we show —through an explicit theoretic analysis on minimal information— that such formulations are not identifiable with existing data. Allowing unresolved early-time dynamics to be absorbed into effective, non-zero initial conditions yields an emergent Markovian description that is maximally informative and consistent across species, organs, particle sizes, and exposure levels. When expressed in normalized variables, uptake dynamics collapse onto a universal trajectory governed by a single dimensionless parameter, the systemic excretion capacity, which is generically small under experimental conditions. The resulting scale-free framework reveals systematic power-law dependencies of enrichment and retention times on ambient concentration, particle size, and body mass. Exploiting this structure, we examine the consistency of extrapolations to humans and show that reported organ burdens —particularly in the brain— are quantitatively compatible with inefficient systemic clearance and strong lipid-driven partitioning. At steady state, human tissue concentrations follow a robust approximate cubic scaling with lipid fraction, identifying lipid content as the dominant and mechanistically interpretable determinant of chronic nanoplastic accumulation.
\end{abstract}

\noindent
Micro- and nanoplastics (MNPs) have emerged as a pervasive class of anthropogenic pollutants capable of entering, circulating within, and accumulating throughout living organisms, particularly vertebrates. Over the last five years, advances in analytical chemistry---including pyrolysis gas chromatography–mass spectrometry (py–GC/MS) and Fourier-transform infrared spectroscopy (FTIR)---combined with controlled laboratory exposure studies, have enabled reliable quantitative measurements of MNP burdens in animal tissues
and, critically, in humans. Py–GC/MS analyses now demonstrate that human organs contain polymer burdens at the level of hundreds to thousands of parts per million (ww), including brain, liver, kidney, testis, and major arteries \cite{nihart2025,Liu2024,Hu2024,Leslie2022}. These observations establish that MNPs cross physiological barriers, accumulate over decades, and exhibit pronounced organ-specific partitioning whose mechanistic basis remains poorly understood.

Parallel short-term studies in model vertebrates have begun to characterize the kinetics of nanoplastics (NPs) uptake and clearance under controlled exposures. Using polystyrene nanoplastics (PS NPs) in zebrafish (\textit{Danio rerio}), Habumugisha~\textit{et~al.}~\cite{habumugisha2023,habumugisha2025} quantified time-resolved concentrations in multiple organs following exposures in the range 1--15\,ppm using MALDI–TOF-MS. Zhang~\textit{et~al.}~\cite{Zhang2025} reported analogous results for protein-coated PS NPs using fluorescence spectroscopy. Earlier work examined microplastic uptake in red tilapia (\textit{Oreochromis niloticus}) \cite{Ding2018} and medaka (\textit{Oryzia melastigma}) \cite{ZHENG2024}, while Choi~\textit{et~al.} \cite{Choi2023} used py–GC/MS to resolve sub-day uptake dynamics in \textit{Zacco platybus}. Together, these studies provide the most detailed dataset currently available on early-time NP toxicokinetics, yet no unified mechanistic framework has existed to interpret them across species, organs,
exposure levels, and analytical methods.

To date, fish bioaccumulation datasets have been interpreted almost exclusively using classical one-compartment exponential kinetics. In the present critical review, we reanalyze a consistent set of zebrafish, medaka, and red tilapia datasets \cite{habumugisha2023,habumugisha2025,Zhang2025,ZHENG2024,Ding2018,Choi2023}, spanning nearly three orders of magnitude in adult body mass, within a multi-compartment toxicokinetic framework. The strong organ specificity of long-term burdens $C_T(t\!\to\!\infty)$ suggests that uptake and bioaccumulation can be reduced to a sequential architecture comprising (i) a systemic organism-wide ``gate–blood’’ compartment controlling the blood concentration $C_S(t)$ and (ii) multiple tissue compartments where endothelial glycocalyx transport and local composition determine retention and partitioning
(Fig.~\ref{fig:schematic}).

The modeling framework builds on the established lineage of pharmacokinetics and toxicokinetics, from the foundational treatments of Gibaldi and Perrier \cite{gibaldi1982} and Rowland and Tozer \cite{rowland2011}, through formal compartmental analysis \cite{jacquez1996,cobelli2000tracer}, to aquatic toxicokinetic and food-web models \cite{Nichols1990,Arnot2004,Spacie1982}. However, biological transport of NPs across physiological barriers involves multi-scale interactions—such as protein corona evolution and transient trapping in the endothelial glycocalyx—that can deviate from classical Fickian behavior. While such anomalous dynamics can be formulated using fractional-order toxicokinetics with Mittag–Leffler kernels, we show in the Supplementary Materials that the application of these models is fundamentally limited by a severe \emph{minimal information problem}: systemic concentrations are not directly observable, and available uptake and depuration datasets are sparse, noisy, and insufficient to resolve higher-order models without parameter degeneracy.

Under these constraints, we demonstrate (see Supplementary Materials) that the apparently complex and method-dependent MNP kinetics in teleosts can be unified within a normalized Sequential Two-Compartment toxicokinetic (S2CT) model, in which systemic and tissue compartments are coupled by first-order Markovian transport. By introducing an effective renormalization through non-zero initial conditions, we show that the dynamics collapse onto a universal formulation governed by a single dimensionless control parameter: the normalized systemic excretion capacity $\eta=\lambda_{Se}/\lambda_{Te}$. Remarkably, across six independent experimental studies employing disparate analytical techniques, this parameter is generically small, yielding a robust master curve that collapses all uptake and depuration data. With the sole exception of {\it Z. platypus} ($\eta \simeq 0.33$), all species exhibit negligible systemic excretion capacity ($\eta \to 0$), suggesting that species-specific physiological factors —potentially related to renal architecture, hepatobiliary function, or immune cell activity— regulate nanoparticle clearance under conditions that warrant further investigation.

In its full generality, MNP bioaccumulation depends on many factors, including ambient concentration $C_w$, particle size and shape, polymer composition, organism body weight $W$, organ mass fractions, and organ-specific composition and bio-architecture. However, the existing literature provides sufficiently consistent and comparable information only for three parameters: $C_w$, particle size $d$, and $W$. We therefore restrict the analysis to these experimentally accessible variables, further focusing on a single polymer composition (polystyrene) because it is currently the only one for which sufficiently consistent, time-resolved, multi-organ bioaccumulation data exist to support a generalizable toxicokinetic analysis. Despite this severe reduction of available information, the results of this study show that, within the relative constraints imposed by biological allometry \cite{Valkengoed2025}, these three parameters capture the dominant scaling behavior of bioaccumulation across species and organs. This reduction yields a scale-free framework that is strongly consistent across datasets and potentially enables quantitative extrapolation to humans.

Exploiting this scale-free structure, we extrapolate teleost-derived parameters to humans and rationalize the observed preferential accumulation of NPs in lipid-rich organs such as brain and liver. We find that long-term organ burdens obey a robust approximate cubic lipid law,
\[
C_T(\infty)\propto f_{\mathrm{lipid}}^{3},
\]
yielding blood-to-brain enrichment factors $K_S\sim10^3$--$10^4$, consistent with recent human measurements \cite{nihart2025}. This behavior, which aligns with the well-established pharmacokinetics of lipophilic compounds \cite{gibaldi1982,rowland2011,Poulin2002,Rodgers2006}, is theoretically supported by two orthogonal concomitant effects: (i) an affinity-driven efficiency of lipid attachment to the polymeric particle, and (ii) a geometric confinement limitation among neighboring particles in the parenchyma. 
Within this framework, the brain emerges as a dominant sink for planetary plastic pollution.

Finally, the translational relevance of teleost models—particularly \textit{D. rerio}—is well established in comparative toxicology and pharmacology \cite{Nichols1990,Arnot2004,Lieschke2007,Brunton2007,JEONG2008,
Santoriello2012,Howe2013,MacRae2015}. The toxicokinetic architecture governing blood–tissue exchange is highly conserved across teleosts and mammals, including humans, spanning vascular organization, endothelial transport, and neurovascular interfaces \cite{Katharios2004,Walle2004,Murphey2006,JEONG2008,
Li2011,Hung2012,Luckenbach2014,Diotel2018,Jurisch-Yaksi2020,
Ikeshima-Kataoka2022}. This conservation underpins the quantitative extrapolation developed here and supports the use of a minimal, mechanistically grounded S2CT framework to bridge short-term animal kinetics and long-term human bioaccumulation.

\section*{Mechanistic framework: a multi-organ, sequential two-level (compartment) toxicokinetic model}
\label{sec:framework}

\begin{figure}
    \centering
    \includegraphics[width=0.80\textwidth]{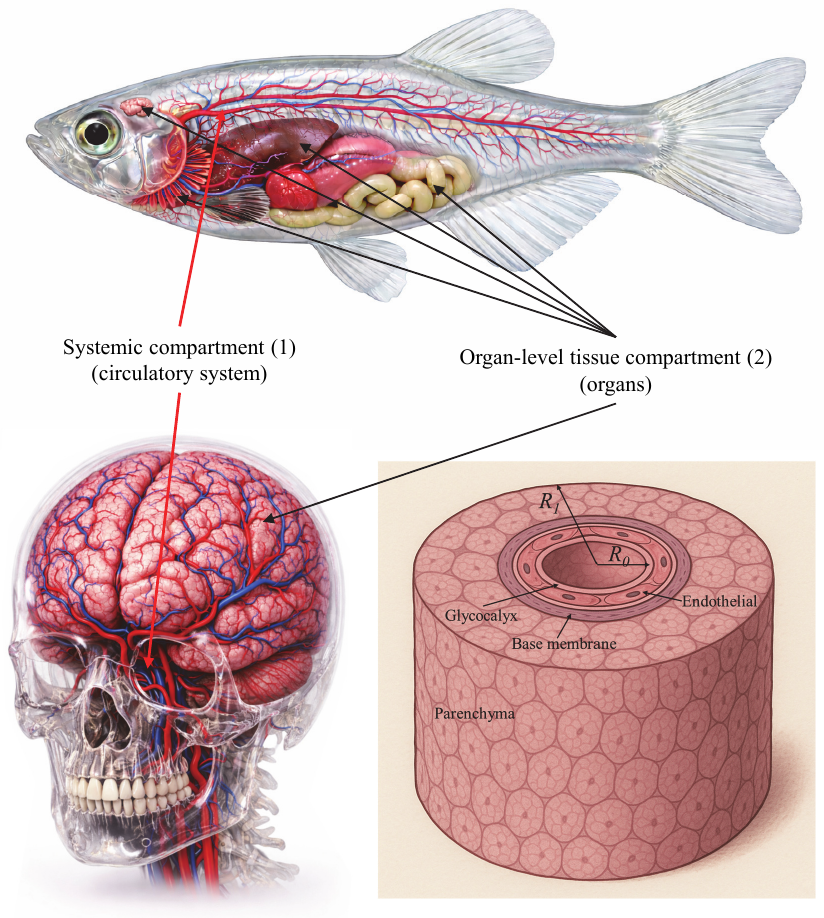}
    \caption{Schematic representation of the sequential two-level toxicokinetic architecture (S2CT model): the systemic compartment (circulatory system, organ-agnostic) linking the external concentration $C_w$ to the homogenized systemic concentration $C_S$ via the three entry gates (respiratory, oral and skin) and rapid circulatory mixing. Organ-level tissue compartment (organs) in which MNPs diffuse (bottom right panel) from the vascular lumen across the endothelial glycocalyx (eGC), endothelium, and basement membrane into the parenchymal domain of radius $R_1$  (e.g. $R_1 \approx 15~\mu$m) surrounding a capillary of radius $R_0$ (e.g. $R_0 \approx 3~\mu$m). Short-term kinetics are dominated by the eGC, while long-term accumulation reflects organ-specific physicochemical composition.}
    \label{fig:schematic}
\end{figure}

The biological transport of micro- and nanoplastics (MNPs) in vertebrates is governed by a hierarchy of barriers and mixing processes that cannot be captured by classical Fickian pharmacokinetics. Motivated by anomalous diffusion theory \cite{bouchaud1990anomalous,Metzler2000,ben-avraham2000diffusion}
and by converging evidence from recent short- and long-term kinetic measurements in fish and humans
\cite{habumugisha2023,habumugisha2025,Zhang2025,Ding2018,ZHENG2024,nihart2025,Leslie2022,Liu2024,Hu2024}, we initially formulate a fractional \emph{sequential two-compartment toxicokinetic} (S2CT) model. The framework isolates (i) a systemic, organism-wide ``gate--blood'' compartment controlling the time dependence and susceptibility of the systemic concentration $C_S(t)$ through oral, respiratory, and dermal interfaces, and (ii) organ-specific tissue compartments where endothelial glycocalyx (eGC) transport and local physicochemical composition determine uptake, retention, and long-term partitioning.
Figure~\ref{fig:schematic} summarizes the resulting two-level architecture.

The first compartment represents the vascular lumen of the entire circulatory system, acting as the primary distributor of NPs that have crossed the external entry barriers. It is modeled as a well-mixed, capacitive reservoir with spatially uniform concentration $C_S(t)$. This approximation is justified by the large separation of timescales: blood mixing occurs over minutes, whereas uptake and depuration processes occur over days to weeks. As a result, any particle entering the circulation rapidly presents a uniform driving concentration to all downstream organs.

The systemic kinetics comprise a source term describing net uptake from the environment at concentration $C_w$, regulated by the rate $\lambda_{Su}$, and a clearance term characterized by $\lambda_{Se}$. The latter aggregates multiple removal pathways, including sequestration by the reticuloendothelial system, uptake by circulating immune cells, and transfer to tissues. Although individual processes may initially exhibit non-Fickian dynamics (see Suppl. Info.), available measurements indicate that an effective linear description rapidly emerges. Accordingly, the systemic compartment is described by
\begin{equation}
    \frac{d^{\alpha_S}C_S}{dt^{\alpha_S}} = \lambda_{Su} C_w - \lambda_{Se} C_S(t).
\label{eq:fPK1}
\end{equation}
where $d^{\alpha}/dt^{\alpha}$ denotes the Caputo fractional derivative of order $\alpha$ (see Supplementary Materials).

Downstream of the glycocalyx, the endothelial cytoplasm, basement membrane, and adjacent parenchymal interstitium can likewise be treated as a second capacitive system. Within these layers, transport involves hindered diffusion, reversible adsorption, and transient storage, whose characteristic times exceed the diffusion time across individual micro-layers. Consequently, the composite barrier may be coarse-grained into a single lumped compartment with effective uptake and excretion rates.

At the tissue scale, we consider the parenchyma surrounding a representative capillary as a radially symmetric annulus $R_0<r<R_1$, with a no-flux boundary condition at $r=R_1$. Under the capacitive assumptions for the luminal and extraluminal domains and anomalous transport concentrated at the endothelial interface, the annulus can be homogenized into a zero-dimensional compartment. Absorbing geometric factors into the transport coefficients yields
\begin{equation}
(R_1^2-R_0^2) \frac{d^{\alpha_T}C_T}{dt^{\alpha_T}}
=  R_0 \left( P_{u}\, C_S(t) - P_{e}\, C_T(t) \right)
\Longrightarrow
\frac{d^{\alpha_T}C_T}{dt^{\alpha_T}}
= \lambda_{Tu}\, C_S(t) - \lambda_{Te}\, C_T(t),
\label{eq:fPK2}
\end{equation}
with transport coefficients $P_{u,e}$ and
\begin{equation}
\lambda_{Tu,Te} =
P_{u,e}\,\frac{R_0}{R_1^2 - R_0^2}
= P_{u,e} \left(R_0\left(1/V_c-1\right)\right)^{-1},
\label{eq:fpk_core}
\end{equation}
where $V_c=(R_0/R_1)^2$ is the vascular volume fraction. This geometric parameter already confers organ specificity to the model by setting the relative capacity of the systemic compartment. Further bio-functional analysis reveals that $V_c$ correlates with organ protein fraction and is inversely correlated with lipid fraction (see Supplementary Materials), reflecting strong multi-correlation among vascularization, protein content, and lipid content across organs. However, among these interdependent variables, only lipid fraction exhibits a robust and mechanistically explainable correlation with chronic tissue concentration, consistent with the hydrophobic nature of polystyrene and its preferential partitioning into lipid-rich environments. Within this reduced description, both compartments behave as capacitive reservoirs coupled through an anomaluos resistor associated with the eGC–endothelium complex.

A fully general description of micro- and nanoplastic transport can be formulated using fractional-order toxicokinetic equations with memory kernels, as detailed in the Supplementary Materials (Supplementary Materials). Analysis of the available datasets, however, shows that this formulation is underdetermined, reflecting a minimal information problem. We therefore allow an initial non-Markovian transient to elapse and describe the experimentally resolved regime using effective Markovian dynamics. The unresolved early-time complexity is absorbed into unknown, non-zero effective initial conditions for the systemic and tissue compartments, subsequently used to renormalize the observable variables (canonical reduction shown in next section). Hence, System~(\ref{eq:fPK1})--(\ref{eq:fPK2}), with $\alpha_S = \alpha_T = 1$ and initial conditions $C_{(S,T)}(t_0)=C_{(S,T)0}$, admits the classical solution~\cite{jacquez1996}
\begin{align}
    C_S(t) & =C_{S\infty}+(C_{S0}-C_{S\infty})\mathrm{e}^{-\lambda_{Se}t},
    \label{eq:solution_S}\\
    C_T(t) & =C_{T\infty}+(C_{T0}-C_{T\infty})\mathrm{e}^{-\lambda_{Te}t}
    + \alpha (\mathrm{e}^{-\lambda_{Te}t}-\mathrm{e}^{-\lambda_{Se}t}),
    \label{eq:solution_T}
\end{align}
with
\begin{equation}
    C_{S\infty}=K_S C_w \equiv
    \frac{\lambda_{Se}}{\lambda_{Su}} C_w,\quad C_{T\infty} = \frac{\lambda_{Tu}}{\lambda_{Te}} C_{S\infty} = \frac{\lambda_{Tu}}{\lambda_{Te}} \frac{\lambda_{Su}}{\lambda_{Se}} C_w \equiv K_T K_S C_w
\end{equation}
and
\begin{equation}
    \alpha= K_T \frac{C_{S0}-C_{S\infty}}{1-\eta},
    \qquad
    \eta=\frac{\lambda_{Se}}{\lambda_{Te}}.
\end{equation}
The dimensionless ratio $\eta$ defines the \emph{excretion capacity} of the tissue via the systemic compartment.

During uptake or depuration, the external forcing $C_w$ is finite or vanishing, respectively, yielding finite or zero steady states. In the depuration regime, $C_w=0$ implies $C_{(S,T)\infty}=0$. We therefore distinguish explicitly between uptake and depuration solutions, denoted $y_{(S,T)u}$ and $y_{(S,T)d}$.

\subsection*{Canonical reduction of the toxicokinetic uptake model}

We consider uptake processes initiating from effectively vanishing concentrations and asymptotically approaching steady states. Early-time dynamics may involve unresolved multiscale and non-Markovian effects; here, these are absorbed into unknown initial conditions defined at the onset of effective Markovian behavior. Exploiting this freedom, normalization by $C_{(S,T)0}$ yields a universal canonical form in which the entire family of uptake solutions depends on the single parameter $\eta$.

Dividing Eqs.~(\ref{eq:solution_S})--(\ref{eq:solution_T}) by the initial
conditions and introducing
\begin{align}
x & =\lambda_{Te}t,\\
\xi_T & \equiv \frac{C_{T\infty}}{C_{T0}}=\eta^{-1},\\
\xi_S & \equiv \frac{C_{S\infty}}{C_{S0}}=\eta^{-1}(3-2\eta)^{-1},
\end{align}
one obtains
\begin{align}
    y_{Su} \equiv \frac{C_S}{C_{S0}} & = \xi_S+(1-\xi_S)\mathrm{e}^{-\eta x}, \\
    y_{Tu} \equiv \frac{C_T}{C_{T0}} & = \xi_T-\mathrm{e}^{-x}+(2-\xi_T)\mathrm{e}^{-\eta x}.
    \label{eq:canonical_T}
\end{align}
In the limit $\eta \to 0$, the tissue uptake kernel reduces to
\begin{equation}
    \left.y_{Tu}\right|_{\eta \to 0}=2+x-\mathrm{e}^{-x}.
    \label{eq:asympto_u}
\end{equation}

\subsection*{Canonical toxicokinetic depuration model}

In contrast to uptake, the depuration phase is defined by the fixed time at which external forcing is discontinued. The initial conditions are therefore fixed \emph{a priori} and cannot be absorbed into a renormalization, requiring an additional dimensionless parameter.
The normalized depuration solutions read
\begin{align}
    y_{Sd} & =\mathrm{e}^{-\eta x},\\
    y_{Td} & =\mathrm{e}^{-x}(1-\xi_d)+\xi_d \mathrm{e}^{-\eta x},
    \label{eq:depuration}
\end{align}
with
\begin{equation}
    \xi_d=\frac{\lambda_{Tu}}{\lambda_{Te}}
    \frac{C_S(t_p)}{C_T(t_p)}(1-\eta)^{-1}
    \equiv k_d (1-\eta)^{-1}.
\end{equation}
The uptake and excretion rates generally differ, reflecting the irreversibility
of biological handling of foreign materials.
In the limit $\eta \to 1$,
\begin{equation}
    \left. y_{Td}\right|_{\eta \to 1} =
    \mathrm{e}^{-x}\left(1 + k_d x \right).
\end{equation}

In the uptake regime, physically meaningful solutions require $\eta<1$, whereas depuration solutions remain well defined for all $\eta$. Figure~\ref{fig:y_Tu} in the Supplemental Materials illustrates the normalized kernels. Notably, the tissue concentration may initially increase even while the systemic concentration decreases, reflecting a transient positive flux from the systemic to the tissue compartment. This effect is enhanced when systemic clearance exceeds tissue elimination ($\eta>1$).

Having established the exact analytical structure of the Markovian S2CT model, we next apply it to experimental datasets to extract kinetic parameters and assess its predictive power across species, organs, and exposure regimes.


\section*{Analysis of experimental data and results}

The analysis of the selected experimental measurements comprises two distinct frameworks:

\begin{enumerate}
\item Short-term kinetic studies \cite{habumugisha2023,habumugisha2025,Zhang2025,Ding2018,ZHENG2024}, which investigate the kinetics of polystyrene nanoparticles (PS NPs) in {\it four} teleosts species (\textit{Danio rerio, Oreochromis niloticus, Oryzias melastigma} and {\it Zacco platybus}). This first group of datasets corresponds to \emph{short-term exposures} (from 1 up to 46 days). These studies quantify the uptake, bioaccumulation, and excretion of PS NPs across different tissues under controlled laboratory conditions, using spectroscopic or mass-spectrometric techniques (discussed in section Methods). Their data provide time-resolved concentration profiles suitable for fitting with the analytical solutions to extract the essential transport coefficient $\lambda_{Te}$ and the normalized enrichment factor.

\item Long-term studies \cite{nihart2025,Liu2024,Hu2024}, which examine the steady-state bioaccumulation of nanoplastics (NPs) of mixed polymer composition across multiple human organs measured across multiple human organs over several periods (2016 to 2024). These datasets reveal organ-specific enrichment factors that depend systematically on biochemical and biophysical organ parameters—most notably the lipid content, which correlates with lipid fraction and, indirectly -or consequently-, to vascular volume. The objective of the long-term analysis is thus to quantify these organ-specific partition coefficients and to relate them to tissue composition and the progressive environmental rise in ambient NP concentration over time.
\end{enumerate}

The following subsections will address separately both frameworks, with the ultimate goal to harmonize the kinetic findings with the long term bioaccumulation burdens observed in humans.

\subsection*{Comparative assessment of PS NP temporal bioaccumulation datasets and implications for kinetic modeling}

During acute or subchronic exposures, the dominant process is the {rapid partitioning of NPs from the external medium} (water, air, ingested food) {into the systemic circulation} via the principal exposure gates: the respiratory epithelium (gills in fish, alveoli in mammals), the gastrointestinal mucosa, and---to a lesser extent---the skin. These barriers are characterized by high surface area, fenestrated or discontinuous capillaries (in liver sinusoids and intestinal villi), and active transcytotic machinery optimized for nutrient absorption. Once particles are internalized and stabilized by endogenous protein coronas (albumin, apolipoproteins, immunoglobulins...), they enter a well-mixed blood compartment with a characteristic circulation time of order 1 minute in vertebrates. In this regime, the {systemic concentration} $C_{S\infty} = K_S C_w$ is determined almost entirely by gate-level processes: mucosal permeability, corona-mediated adhesion, and systemic clearance via hepatobiliary and renal routes. All downstream tissues --brain, liver, muscle, kidney-- experience the same circulating $C_S$, and the differences in tissue burden arise primarily from {biological composition and geometric factors -the vascular volume fraction $V_c$}-, leading to different effective rates $\lambda_{T(u,e)}$.

\begin{figure}
\centering
\includegraphics[width=0.60\textwidth]{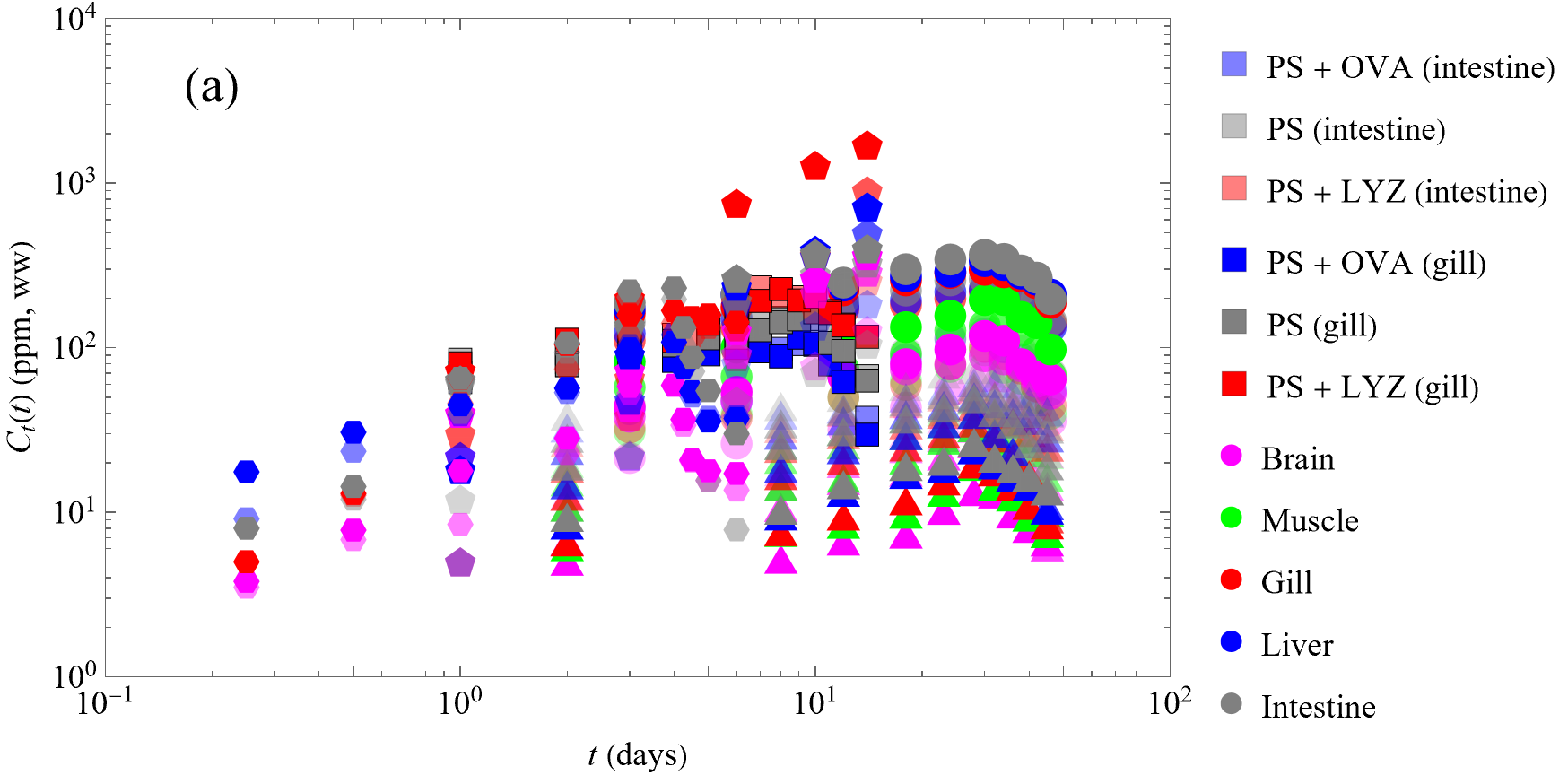}\\
\includegraphics[width=0.59\textwidth]{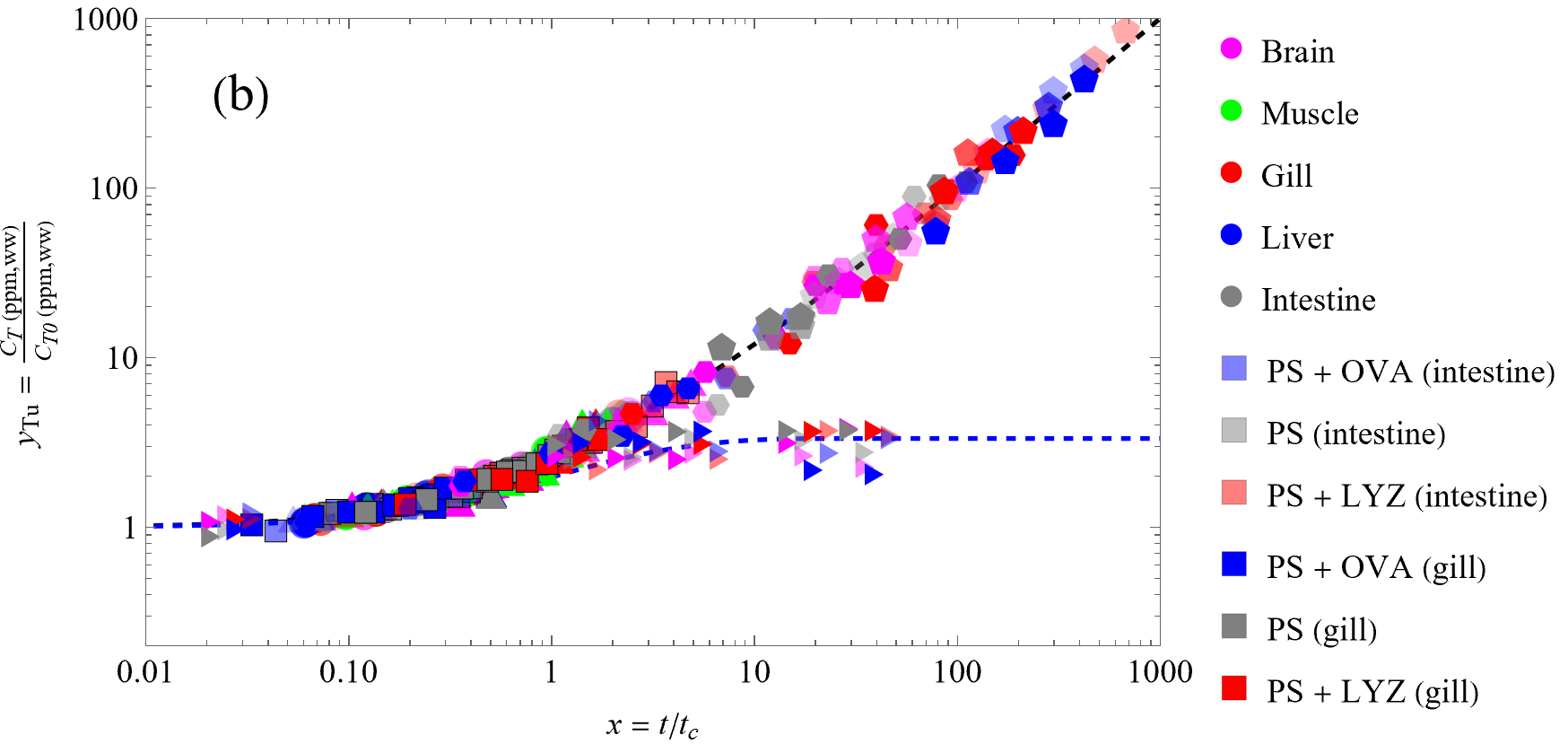}\includegraphics[width=0.41\textwidth]{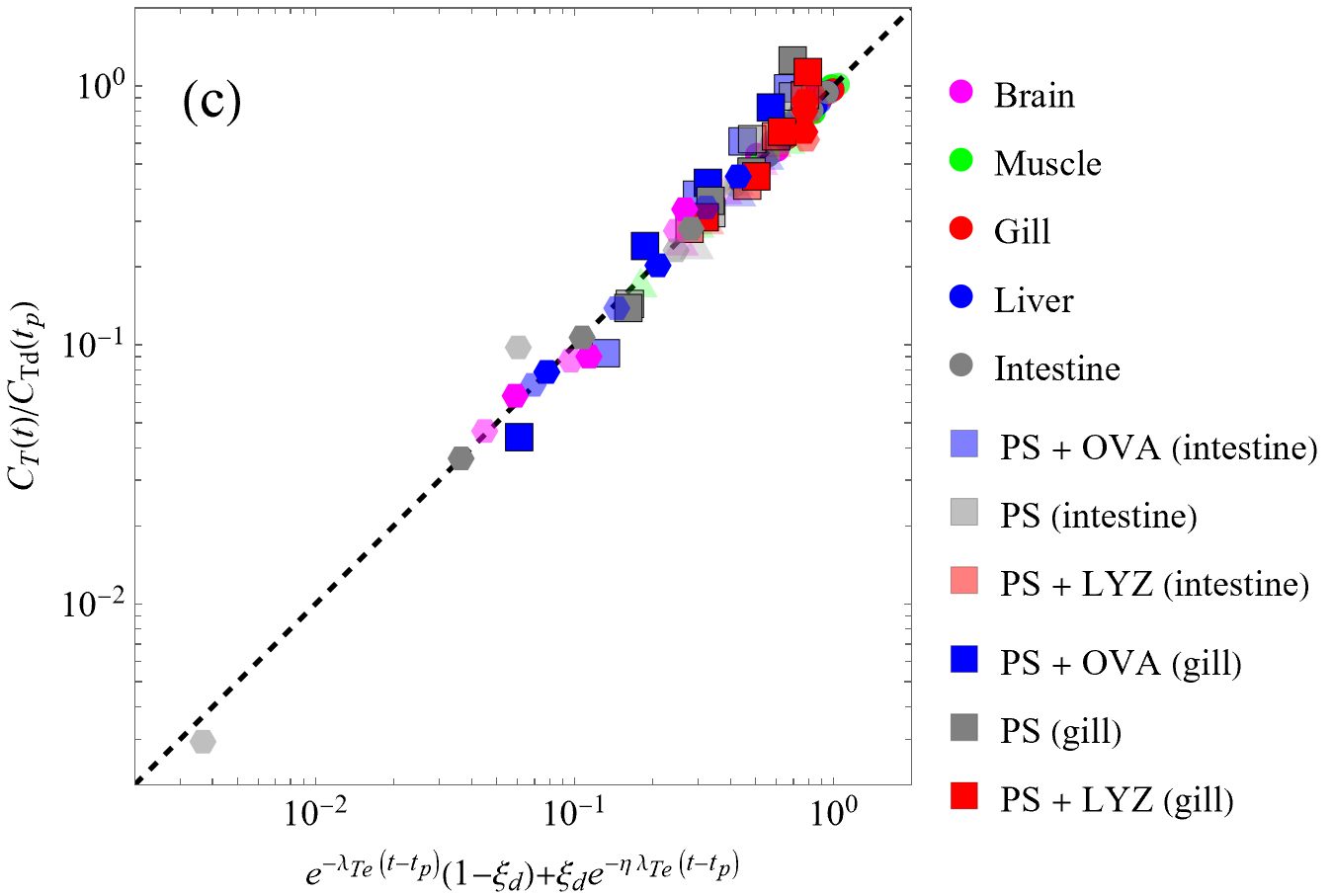}\\
\caption{(a) Raw data (up to five organs) from the six studies selected \cite{habumugisha2023,habumugisha2025,Zhang2025,ZHENG2024,Ding2018,Choi2023} on NP bioaccumulation process in teleosts (400 measurements). Color codes for each organ are indicated. The opacity of colors indicate: (1) Circles, opacity proportional to ambient concentrations $C_w={5,10,\text{ and }15}$ ppm (study of 2023 \cite{habumugisha2023}, {\it Danio rerio}); (2) Triangles, opacity proportional to particle size (study of 2025 \cite{habumugisha2025} {\it Danio rerio}), ambient concentration 1 ppm; (3) Squares: the selected data from \cite{Zhang2025} ({\it Danio rerio}, intestine and gill), ambient concentration 0.45 ppm, $\sim$525 nm PS NPs, with and without protein coronas (ovoalbumin and lysozyme); (4) Pentagons: data (four organs) from \cite{Ding2018} ({\it Oreochromis niloticus}), opacity proportional to logarithm of water concentrations of NPs $C_w={0.001, 0.01, \text{ and }0.1}$ ppm. (5) Hexagons: data (four organs) from \cite{ZHENG2024} ({\it Oryzia melastigma}) excluding eye and skin for consistency with the rest of data, PS NPs of 120 nm (not their 2 $\mu$m data), and two water environments (fresh and synthetic seawater). (6) Tilted triangles: data (four organs) from \cite{Choi2023} ({\it Zacco platybus}), PS NPs of 50 nm, two water concentrations (0.01 and 0.1 ppm).  The logarithmic abscissa emphasize the initial short-term dependencies. (b) Universal best fitting of the six uptake data series considered. Black and blue dashed lines are function $y_{Tu}$ given by (\ref{eq:canonical_T}) for $\eta=0$ (asymptotic form) and $0.33$, respectively. (c) Universal best fitting of the {\it four} depuration data series available from the considered studies on NP bioaccumulation processes \cite{habumugisha2023,habumugisha2025,ZHENG2024,Zhang2025}. Black dashed line is the identity.}
\label{fig:raw}
\end{figure}

Figure \ref{fig:raw}(a) gives an illustrative view of the raw data used in this study. We digitized PS-NP bioaccumulation time courses from six selected experimental studies \cite{habumugisha2023,habumugisha2025,Zhang2025,ZHENG2024,Ding2018,Choi2023}:

(i) Habumugisha {\it et al.} \cite{habumugisha2023,habumugisha2025} across five tissues (intestine, liver, gill, muscle, brain) of {\it Danio rerio}. Collectively, these two studies span five particle sizes (20, 50, 100, 200 and 500 nm), and four exposure concentrations (1, 5, 10, and 15 ppm in the ambient fresh water). However, while the first study \cite{habumugisha2023} describes the exposure to 50 nm PS NPs at three concentrations (5, 10 and 15 ppm), the second describes it for an ambient concentration of 1 ppm in the case of four PS NPs sizes (20, 100, 200 and 500 nm). The analytical method used in this study is MALDI-TOF-MS.

(ii) Zhang {\it et al.} \cite{Zhang2025}, selecting data from the intestine and gills of {\it Danio rerio} (0.35 $\pm$ 0.046 g) exposed to 0.46 ppm of $\sim$ 525 nm PS NPs in fresh water in three conditions: (1) pristine particles, and particles coronated by (2) ovoalbumin, and (3) lysozyme. This study uses fluorescence (fluorescently labeled polystyrene nanoparticles).

(iii) Zheng {\it et al.} \cite{ZHENG2024}, across four tissues (intestine, liver, gill, brain) of {\it Oryzias melastigma} (estimated at 25 mg $\pm$ 8 mg), using two particle sizes (120 nm and 2 $\mu$m) and two water environments (fresh and synthetic seawater) allowed by this species. They use photoluminiscence spectrometry (fluorescence).

(iv) Ding {\it et al.} \cite{Ding2018}, across the same previous four tissues of {\it Oreochomis niloticus} (21 $\pm$ 3.9 g), using three {\it low concentrations} (0.001, 0.01 and 0.1 ppm) of 100 nm fluorescent PS NPs in fresh water. They use the same method of Zheng {\it et al.}.

(v) Choi {\it et al.} \cite{Choi2023}, across the same previous four tissues of {\it Zacco platybus} (1.402 $\pm$ 0.03 g), using two {\it low concentrations} (0.01 and 0.1 ppm) of 50 nm fluorescent PS NPs in fresh water. They use Py-GC/MS spectrometry.

Details of the experimental methodologies and associated uncertainties of these studies are provided in the Methods section.

\subsection*{Canonical uptake fitting}


The canonical representation of nanoparticle uptake dynamics was systematically examined across the selected experimental datasets. For each study, the uptake time series were fitted in logarithmic least-squares to the canonical solutions~(\ref{eq:canonical_T}), yielding best quantitative estimates of the tissue excretion rate $\lambda_{Te}$ and the initial uptake enrichment factor $C_{T0}/C_w$. A central outcome of this analysis is that, with the sole exception of the dataset reported by Choi \textit{et al.}~\cite{Choi2023}, the fitted uptake curves display no detectable sensitivity to the value of the excretion capacity $\eta$. Consequently, for all remaining datasets we adopt the asymptotic formulation~(\ref{eq:asympto_u}), which is valid in the limit of small $\eta$. This result indicates that, under the experimental conditions considered, the effective excretion capacity of the organisms is sufficiently small to lie below the resolution of the available data.

\paragraph{Singularity of {\it Z. platibus}?} The Choi \textit{et~al.} time series is optimally described by $\eta \simeq 0.33 \pm 0.01$, confirming that this experiment probes a dynamical regime beyond the strict asymptotic limit captured by $\eta = 0$ (i.e., time scales smaller than $\lambda_{Se}^{-1}$). This behavior identifies the Choi \textit{et~al.} dataset as a meaningful outlier and indicates that, for this specific study, the excretion dynamics of the species involved (\textit{Z.~platypus}) differ quantitatively from those observed in the other experimental systems. The authors specifically indicate that their measurements point to a common excretion mechanism shared by other study involving coronated gold nanoparticles \cite{Zhu2010}. Importantly, this deviation cannot be attributed to the use of the Py-GC/MS quantification methodology, which—if anything—provides the most chemically specific and least biased estimate of polymer mass. Rather, the need for $\eta \sim \mathcal{O}(1)$ suggests that the balance between uptake and clearance in \textit{Z.~platypus} operates on a timescale comparable to the experimental observation window of other species, potentially placing this species outside the strictly asymptotic excretion regime sampled by the other studies, at least under the experimental conditions considered in their study. In this sense, the apparent discrepancy is not a methodological artifact but should reflect a genuine toxicokinetic regime difference which the S2CT framework captures without loss of internal consistency. Until resolved, the {\it Z. platypus} result suggests a distinct kinetic {\it regime} rather than a distinct species {\it phenotype}. This strongly motivates further comparative studies across species and exposure regimes. The final resulting data collapse is shown in Fig. \ref{fig:raw}(b).


As a final validation step, the internal consistency of the fitting procedure is assessed by examining the systematic dependence of the fitted parameters $k_p=C_{T0}/C_w$ and $t_c=\lambda_{Te}^{-1}$ on the defining characteristics of each experimental series, including the ambient nanoplastic concentration $C_w$, the PS~NP diameter $d$, and the average body mass $W$ of the exposed organisms. To this end, we employ the \texttt{LinearModelFit} routine of \textit{Mathematica} on the logarithmic variables, fitting the data to a product-of-powers scaling law of the form $\varphi = A \, C_w^{\phi} d^{\zeta} W^{\omega}$, as commonly used in dimensional analysis and physical scaling arguments. The prefactor $A$ ensures the correct dimensional consistency of the dependent variable $\varphi$ (i.e.\ the generic enrichment factor $k_p$ or the characteristic time $t_c$), while the exponents $\phi$, $\xi$, and $\omega$ encode the corresponding scaling dependencies. The results, summarized in Table~\ref{tab:uptake} (Supplementary Materials), reveal a high degree of internal coherence: the inferred scaling relations yield average Pearson correlation coefficients $R^2$ exceeding $0.91$ for the enrichment factor $k_p$ and $0.79$ for the characteristic time $t_c$ (see Supplementary Materials for further information). This strong and systematic parameter dependence not only provides independent, quantitative support for the proposed toxicokinetic framework; it further suggests that the fitted parameters encode physically meaningful and biologically interpretable trends, rather than incidental numerical correlations, which warrant deeper investigation beyond the scope of this critical review work.

\subsection*{Depuration fitting}

The comprehensive analysis of the uptake dynamics indicates that, under the experimental conditions considered, the excretion rate in the systemic compartment, $\lambda_{Se}$, must be several orders of magnitude smaller than the effective uptake rate of the tissue compartment, $\lambda_{Tu}$. Importantly, however, the depuration process operates in the opposite kinetic direction and should not be interpreted as a simple time reversal of uptake. On the contrary, the distinct kinetic signatures observed experimentally suggest that biological depuration is governed by mechanisms that are both faster and more efficient than those responsible for uptake, at least in healthy organisms. This asymmetry is consistent with the presence of active physiological processes dedicated to the elimination of non-metabolizable or potentially toxic materials. From a kinetic standpoint, this endows living organisms with a rectifying behavior, whereby accumulation and elimination are controlled by fundamentally different rate-limiting steps. As a consequence, the excretion rates need not mirror the uptake rates, and both large and small values of the excretion capacity $\eta=\lambda_{Se}/\lambda_{Te}$ remain physically admissible in this phase.

At present, no experimental data are available that isolate a purely \emph{depuration} phase for the species investigated by Choi \textit{et~al.}, nor for other studies involving the same species. As a result, it is not possible to perform a direct comparison between uptake and depuration kinetics in organisms that exhibit a high apparent excretion capacity already during the uptake phase. In contrast, datasets that provide reliable measurements of both the uptake process and the subsequent depuration phase are available in Refs.~\cite{habumugisha2023,habumugisha2025,Zhang2025,ZHENG2024}, and these form the basis of the present depuration analysis.

For these studies, we compare the experimentally measured values of the left-hand side of Eq.~(\ref{eq:depuration}) with the corresponding theoretical prediction given by the right-hand side, which depends on three fitting parameters: the tissue excretion rate $\lambda_{Te}$, the excretion capacity $\eta$, and the dimensionless parameter group $\xi_d$ (where both $\lambda_{Te}$ and $\eta$ are different from the uptake values). For the purposes of data analysis, $\xi_d$ is treated as a single effective parameter encapsulating the relative initial conditions at the onset of depuration. A least-squares fit of the experimental data to this theoretical expression is shown in Fig.~\ref{fig:raw}(c), yielding best-estimate values of $\lambda_{Te}$, $\eta$, and $\xi_d$ for each dataset (see Supplementary Materials).


Again, the collapse of all depuration datasets onto the proposed S2CT model is tight and provides strong support for the existence of a universal, Fickian mechanism of exponential excretion in teleosts. This result is consistent with earlier studies, which inferred exponential depuration kinetics under the simplifying assumption of a single-compartment Markovian system \cite{habumugisha2023,habumugisha2025,Zhang2025,ZHENG2024}. Importantly, the present framework recovers the same Fickian depuration law within a more refined two-compartment description, in which exponential clearance emerges naturally from a differentiated double-compartment depuration kinetics rather than from a single effective clearance process.

In this study, motivated by the analysis of NP bioaccumulation burdens observed in humans —particularly in the brain— we have focused primarily on the uptake process, which dominates long-term tissue accumulation. A more detailed and quantitative investigation of depuration dynamics is deferred to future work, as additional time-resolved depuration datasets and systematic analyses become available in the literature. Thus, we turn next to the extrapolation of the uptake mechanisms identified in teleost models to human micro- and nanoplastic bioaccumulation, with particular emphasis on accumulation in the human brain.

\section*{From teleosts to humans}

Transforming the raw uptake data into the nondimensional variables $C_T/(k_p C_w)$ and $t/t_c$ yields a tight collapse across highly heterogeneous experimental conditions, supporting the internal consistency of the compiled dataset. The data span four teleost species with body masses differing by nearly three orders of magnitude ($0.02$~g medaka to $21$~g red tilapia), six nanoparticle diameters ($20$--$500$~nm), and seven ambient exposure concentrations covering almost four orders of magnitude ($0.001$--$15$~ppm).

In dimensional form, these factors produce nearly three decades of variation in absolute tissue burdens and wide apparent differences in uptake kinetics (Fig.~\ref{fig:raw}). Yet, after rescaling only by the fitted uptake coefficient $k_p$ and the characteristic time $t_c=\lambda_{Te}^{-1}$, all datasets that operate in the low systemic-excretion regime (i.e., except {\it Z. platybus}) collapse onto a single master trajectory (Fig.~\ref{fig:raw}b). The remaining scatter is small relative to the spread in the raw data and suggests no systematic dependence on species, particle size, or exposure concentration amongst teleosts.

This naturally motivates the extrapolation problem addressed next. In principle, higher vertebrates---including humans---could operate closer to the \textit{Z.~platypus} regime, with comparatively efficient systemic excretion. However, the organ burdens reported by Nihart \textit{et~al.} instead indicate markedly inefficient clearance across all examined organs, including liver, kidney, and brain. The magnitude of brain accumulation, in particular, suggests that under current exposure conditions humans are closer to the low-excretion teleost regime (i.e. studies on {\it D. rerio, O. melastigma} and {\it O. niloticus}) than to a high-clearance scenario ({\it Z. platybus}). We therefore examine whether the uptake mechanisms inferred from low-excretion teleosts can be extrapolated to human NP bioaccumulation, with emphasis on the brain.

\subsection*{Parametrical scenario for humans}

To extrapolate the low-excretion toxicokinetic framework from teleosts to humans, we define a minimal parametric scenario based on:

(i) ambient micro- and nanoplastic exposure concentrations,

(ii) the effective particle-size distribution relevant for human uptake and
tissue translocation, and

(iii) the characteristic body mass of exposed individuals.

\subsubsection*{Ambient concentration}

Large-scale environmental compilations show that microplastic mass concentrations in surface waters and near-surface aquatic environments span orders of magnitude, with logarithmic-average values clustering around $C_{\mathrm{MNP}}\sim10^{-2}\,\mathrm{mg\,m^{-3}}$ \cite{Zhao2025}. These measurements are size-integrated; however, fragmentation-driven size spectra are well represented by scale-free power laws \cite{Collins2025}, implying a finite (but small) mass fraction in the submicron/nanoplastic range. Under conservative assumptions, the mass fraction below $1\,\mu\mathrm{m}$ lies in the range $0.3$--$3\%$ \cite{Zhao2025,Collins2025}, corresponding to an effective ambient
nanoplastic mass concentration $C_{w,<1\mu\mathrm{m}}\sim3\times10^{-8}$--$3\times10^{-7}\,\mathrm{ppm}$. Although minor in mass, this submicron fraction dominates particle number and is the size range most relevant for systemic transport, cellular uptake, and brain exposure.

Combining these ambient levels with realistic human water and air intake rates, a basic mass-balance argument yields a clear inconsistency: even assuming negligible excretion, inhalation and drinking water alone would require unrealistically large cumulative mass fluxes ---equivalently, implausibly long exposure times--- to reach the burdens reported in human brain tissue. This excludes air and water as dominant standalone drivers of the observed bioaccumulation and instead points to diet as the primary intake pathway, consistent with trophic transfer and with the lipid-rich nature of many dietary matrices.

\paragraph{Average dietary concentration implied by human bioaccumulation burdens.}
If submicrometric and nanometric plastic particles are retained with near-unity efficiency ---as suggested by the very small effective systemic excretion rates inferred above--- then the observed organ burdens directly constrain the long-term average plastic concentration in the diet. Taking representative tissue concentrations of order $4\times10^{3}$~ppm (brain), $4\times10^{2}$~ppm (liver), $6\times10^{2}$~ppm (kidneys), $\sim10^{2}$~ppm (skeletal muscle), and $\sim1$~ppm (blood), together with standard adult organ masses, gives a stored plastic mass of order $9$--$10$~g in these major compartments. If the remaining tissues contribute comparably, the total body burden is $\sim2\times10^{1}$~g. If this
mass accumulates over $\sim20$~years with a small excretion capacity, and assuming a representative dietary intake of $\sim2$~kg~day$^{-1}$ (food and beverages, wet mass), the implied mean dietary concentration is $C_{\mathrm{food}}\sim1$--$2$~mg~kg$^{-1}$, i.e.~ $\mathcal{O}(1)$~ppm. This appears consistent with the Py-GC/MS measurements of Jeffries {\it et al.} \cite{Jeffries2025} on beverages. However, seafood, meat or produce may exhibit much larger concentrations \cite{Ribeiro,Leslie2022} of MNPs, a fraction of which is in the NP range. If such exposure is effectively intermittent (during meals), over a daily fraction $\sim10^{-1}$--$10^{-2}$, it yields an effective time-averaged scaling $C_w/C_{\mathrm{food}}\sim10^{-2}$--$10^{-1}$ for a consistent comparison with the constant ambient exposure $C_w$ used in teleost experiments. A direct quantification of size-resolved nanoplastic concentrations in representative diets remains an urgent priority for validating the exposure pathway identified here.

\subsubsection*{Particle size}

Regarding size, the most comprehensive human organ study to date indicates that brain-associated plastic burdens are dominated by nanoplastics rather than micrometer-scale particles \cite{nihart2025}. While Py-GC/MS provides size-agnostic mass quantification, complementary microscopy suggests that brain particles are predominantly submicron and concentrated around $\sim$100--200~nm, with larger microplastics effectively excluded. This size selectivity is consistent with blood--brain barrier transport and implies that the particle population governing brain accumulation lies deep in the NP regime. The relevant study \cite{nihart2025} indicates that nanoparticle diameters of order $10^{2}$~nm are most relevant for human brain uptake, while also emphasizing that a statistically size-resolved treatment will be required to construct a fully integrated bioaccumulation estimate within the proposed multi-parameter framework, as formally developed in the Supplementary Materials.

\subsubsection*{Body mass}

For body mass, we adopt a representative adult human value of $W\simeq70$~kg. Inter-individual variability introduces corrections that are small compared with the orders-of-magnitude ranges spanned by environmental concentrations, particle sizes, and reported tissue burdens. Moreover, because body mass enters the model at the level of the individual rather than the organ, organ burdens for specific individuals can be readily rescaled using the explicitly derived body-mass power-law dependence.

\subsection*{Extrapolating the power laws from teleosts to humans}

Together, the three estimated parameters—ambient exposure concentration, particle size, and body mass—define a baseline scenario against which human uptake dynamics and long-term organ bioaccumulation, particularly in the brain, can be analyzed within the proposed toxicokinetic framework.

\paragraph{An important figure: average observed blood concentration of plastic particles.}
Having estimated representative ambient values of $C_w\sim 10^{-1}$ to $10^{-2}$~ppm for the nanoparticle sizes most relevant for systemic transport, a key benchmark for assessing the effective excretion capacity of the human systemic compartment is the polymer mass concentration measured directly in human blood. In a seminal biomonitoring study, Leslie \textit{et~al.}~\cite{Leslie2022} detected plastic particles in 17 of 22 individuals and reported a mean total quantifiable polymer concentration of $\sim1.6~\mu\mathrm{g\,mL^{-1}}$
($1.6$~ppm). This value is in the range from $10^{1}$ to $10^{2}$ times the long-term $C_w$ values from average dietary concentrations inferred above, indicating a net enrichment between intake and circulation $\lambda_{Te}/\lambda_{Tu}\sim\mathcal{O}(10^1)$ to $\mathcal{O}(10^2)$. Combined with a very limited irreversible systemic clearance, a substantial enrichment and retention occur downstream in tissues, as observed in low-excretion teleost studies at relatively high ambient concentrations \cite{habumugisha2023,habumugisha2025,ZHENG2024,Zhang2025}.

\subsubsection*{Estimated nanoplastic concentrations extrapolated from low-excretion teleost studies and comparison with human data}

\begin{figure}
\centering
\includegraphics[width=0.5\textwidth]{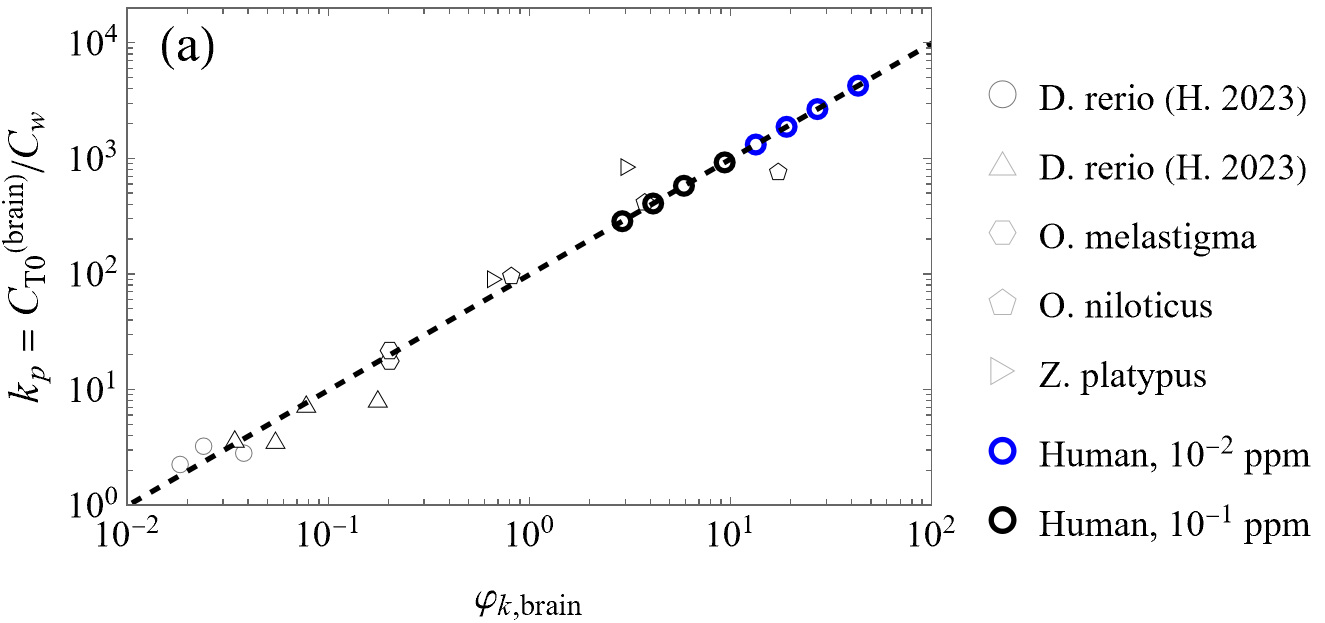}\includegraphics[width=0.5\textwidth]{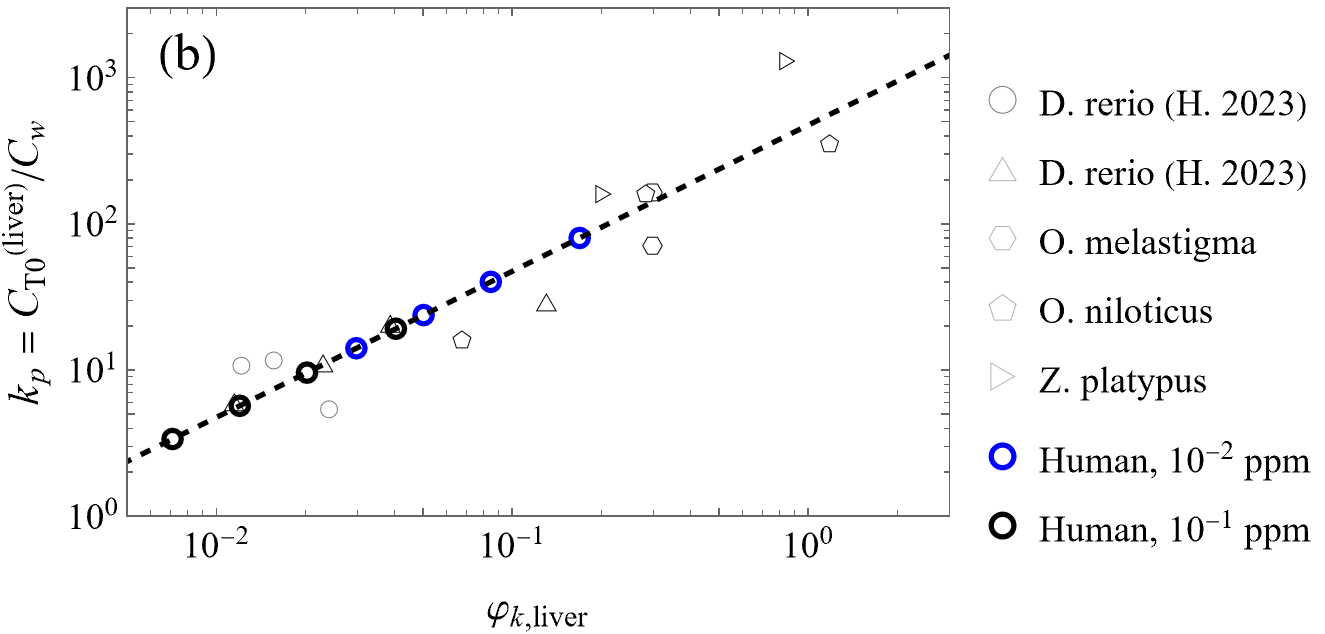}\\
\includegraphics[width=0.5\textwidth]{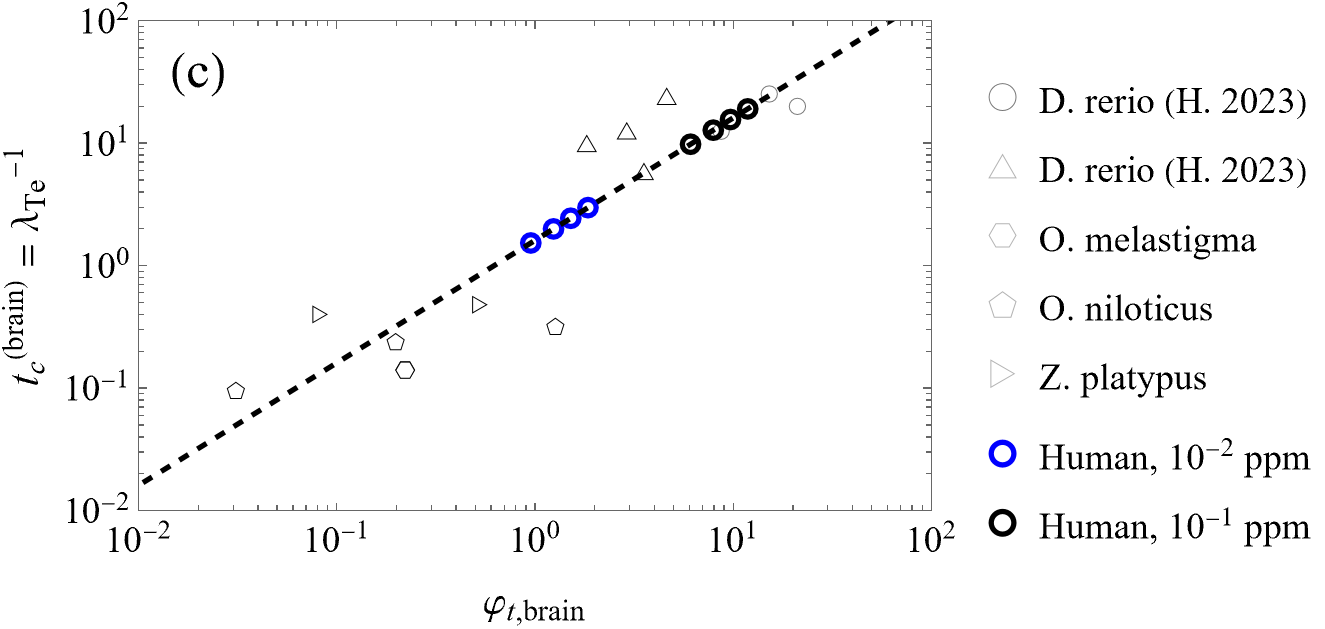}\includegraphics[width=0.5\textwidth]{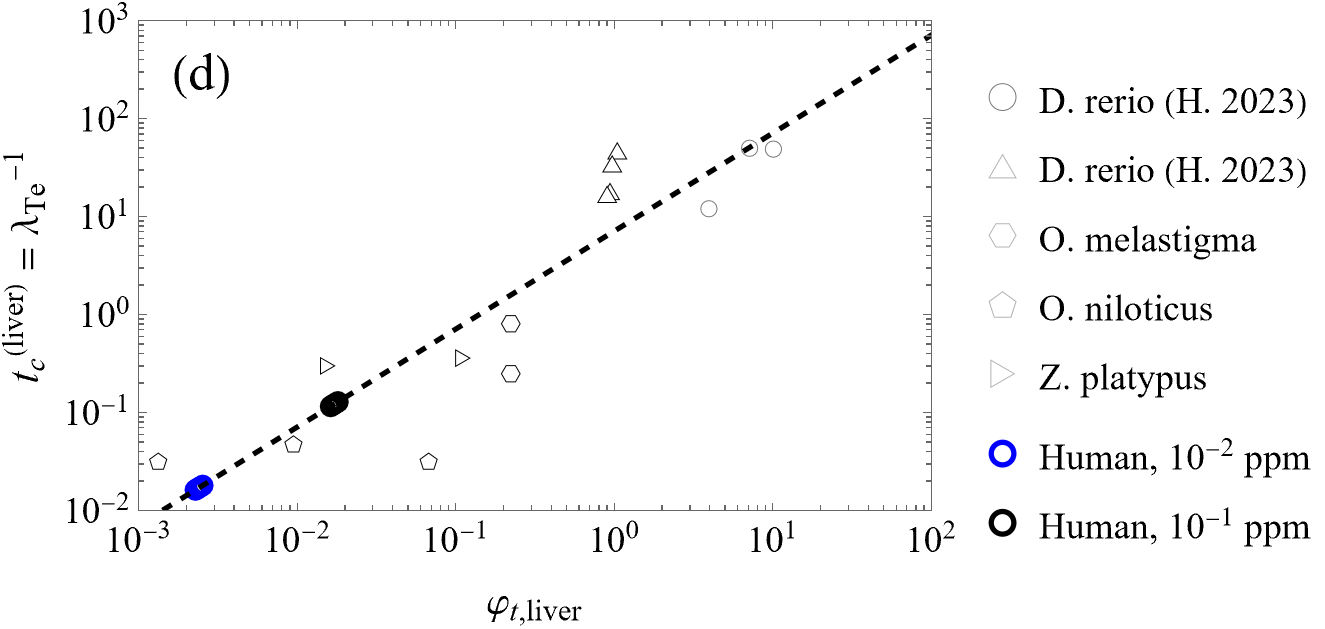}
\caption{Power laws found for coefficients $k_p$ and characteristic times $t_c$, as functions of the variables $\varphi=C_w^\phi d^\zeta W^\omega$ with $\{\phi,\zeta,\omega\}$ powers according to Table \ref{tab:uptake}. Numerical prefactors are consistent with units of length in microns, mass in grams, and time in days. For the human cases, two representative values of {\it average} intake concentrations $C_w$ and for four average values of the particle size $\langle d \rangle=\{20,50,100,200\}$ nm (see Suppl. Info.) are considered.}
\label{fig:extrapolation}
\end{figure}

The power-law scaling relations obtained for the relevant coefficients —tissue concentrations in brain and liver, and the characteristic kinetic rates as functions of ambient concentration, particle size, and body mass— enable a direct extrapolation from teleosts to humans. Placing the human parameter set along these scaling laws and adopting effective ambient exposure concentrations in the range $C_w=10^{-2}$ to $10^{-1}$~ppm yields reference uptake values $C_{T0}$ of roughly $2$--$500$~ppm in the brain and $0.05$--$10$~ppm in the liver. As shown in Fig.~\ref{fig:raw}(b), the ratio of chronic bioaccumulation to initial uptake, $C_{T\infty}/C_{T0}$, spans from order unity under high excretion capacity to values approaching $10^{3}$ in the opposite limit, corresponding to extremely low organism-level clearance ($\eta\sim10^{-3}$ at low ambient concentrations). Data for \textit{Zacco platypus} reported by Choi \textit{et~al.}~\cite{Choi2023} further indicate that effective excretion capacity may increase over time, suggesting adaptive or regulatory responses. Accounting for this behavior, a conservative and biologically plausible range for $C_{T\infty}/C_{T0}$ is $\sim10$--$100$. Within this range, the extrapolated steady-state nanoplastic burdens span approximately $2\times10^{1}$ to $5\times10^{4}$~ppm in the brain and $0.5$ to $10^{3}$~ppm in the liver for particles with diameters below $\sim200$~nm, as summarized in Fig.~\ref{fig:extrapolation}. While these values reflect the broad variability inherent in the extrapolation exercise, they nevertheless provide a meaningful logarithmic-scale estimate that is notably consistent with reported experimental observations \cite{nihart2025}.

Although Nihart \textit{et~al.}~\cite{nihart2025} did not detect polystyrene specifically, they reported substantial burdens of polyethylene and polypropylene in the same organs.
Assuming that polymer hydrophobicity —the dominant physicochemical determinant of tissue interaction and retention— is comparable among these major commodity plastics, the extrapolated values obtained here show not only qualitative but verifiably notable quantitative agreement with the concentrations reported in Fig.~1a,b of Ref.~\cite{nihart2025}.
This concordance, achieved without parameter tuning beyond scaling-law extrapolation, strongly supports the translational validity of the proposed framework from teleosts to humans.

The systematic differences between liver and brain observed both in the teleost-derived scaling laws and in long-term human measurements reveal a robust association between tissue biochemical composition and nanoplastic bioaccumulation. In particular, the enrichment factor $k_p$ scales oppositely in the two organs: liver enrichment decreases weakly with body mass (power-law exponent $\simeq-0.2$), whereas brain enrichment increases with body mass with comparable magnitude but opposite sign. This contrast mirrors the well-established allometric scaling of relative brain size across vertebrates and implicates lipid content as a key determinant of neural accumulation: Although brain mass scales sublinearly with body mass, white matter volume increases disproportionately with brain size, implying an increasing white/gray-matter ratio; because white matter is enriched in myelin, this provides a mechanistic allometric basis for a superlinear increase in effective brain lipid fraction with brain size \cite{Armstrong1983,Zhang2000x}.

\begin{figure}
    \centering
    \includegraphics[width=0.70\textwidth]{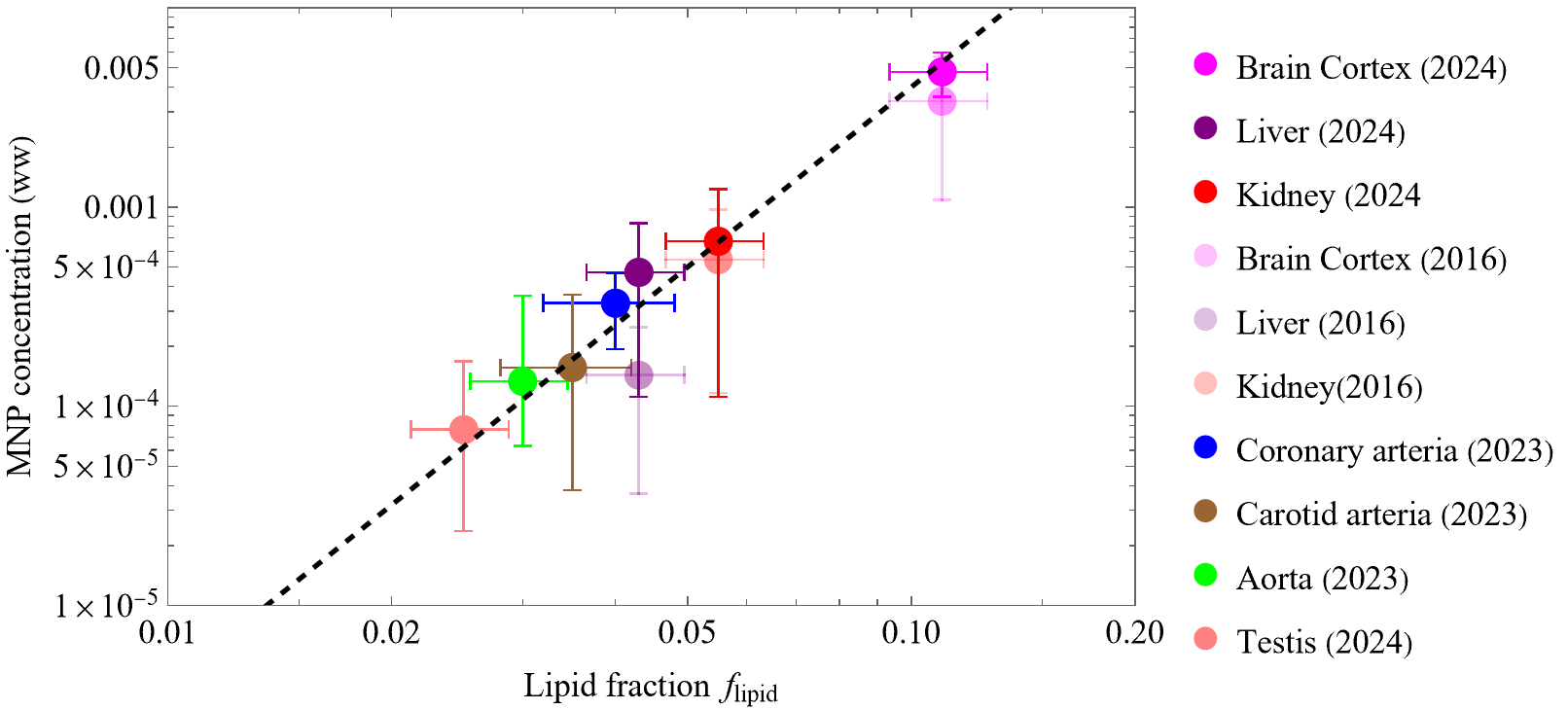}
    \caption{The organ-specific concentrations of micro- and nanoplastics, $C_T$ (wet weight basis), as a function of the lipid fraction $f_{\mathrm{lipid}}$.
    Data compiled from Nihart et al. \cite{nihart2025}, Liu et al. \cite{Liu2024}, and Hu et al. \cite{Hu2024}. Given the relatively small variation of $C_T$ values in the long time scale, compatible with the increment in the concentration of intakes $C_w$ along the years, one can assume $C_T \simeq C_{T\infty}$, the steady state values. The dashed line shows the best-fit power law $C_{T\infty} \propto f_{\mathrm{lipid}}^{3}$, indicating a strong cubic dependence of bioaccumulation on tissue lipid content. Error bars denote the reported experimental uncertainty in both coordinates.}
    \label{fig:MNP_lipids}
\end{figure}

Motivated by this insight, we examined reported bioaccumulation burdens across multiple human organs—including brain, liver, kidney, major arteries, and testis—using available literature data \cite{nihart2025,Liu2024,Hu2024}. Across these tissues, we find a strong correlation between long-term nanoplastic concentration and lipid fraction, summarized in
Fig.~\ref{fig:MNP_lipids}. The data collapse onto a scaling relation approximately cubic in lipid fraction, i.e.
\begin{equation}
C_T(\infty) \propto f_{\mathrm{lipid}}^{\beta},
\end{equation}
with best-fit exponent $\beta \approx 3.0 \pm 0.4$ (95\% CI). While linear partitioning ($\beta$ = 1) is typical for equilibrium lipid-water distribution, the observed superlinear dependence suggests a mechanistic origin of this law, which is theoretically outlined in the Supplementary Materials. The cubic lipid law is neatly obtained by treating the bioaccumulation process as a multiscale phenomenon governed by two orthogonal physical effects taking place in the parenchyma surrounding each NP particle: (i) an affinity-driven efficiency of lipid attachment and (ii) a geometric confinement limitation due to the presence of neighboring particles.
However, given the limited number of organs (n=8) and strong multicollinearity among tissue properties, as detailed in the Supplementary Materials (Fig. S2c-d), this mechanistic interpretation remains tentative pending validation in larger organ sets and across polymer types.


\section*{Acknowledgments}
\paragraph*{Funding:}
Partial funding from the the Spanish Ministry of Science and Innovation, grant no. PID2022-14095OB-C21.
\paragraph*{Competing interests:}
There are no competing interests to declare.
\paragraph*{Data and materials availability:}
Data is recovered from all cited publications and can be available from their authors.


\clearpage 

%

\begin{thebibliography}{10}
\providecommand{\url}[1]{\texttt{#1}}
\expandafter\ifx\csname urlstyle\endcsname\relax
  \providecommand{\doi}[1]{doi:\discretionary{}{}{}#1}\else
  \providecommand{\doi}{doi:\discretionary{}{}{}\begingroup
  \urlstyle{rm}\Url}\fi

\bibitem{nihart2025}
A.~J. Nihart, \emph{et~al.}, Bioaccumulation of microplastics in decedent human
  brains. \emph{Nature Medicine} \textbf{31}, 1114--1119 (2025),
  \doi{10.1038/s41591-024-03453-1}.

\bibitem{Liu2024}
S.~Liu, \emph{et~al.}, Microplastics in three types of human arteries detected
  by pyrolysis-gas chromatography/mass spectrometry (Py-GC/MS). \emph{J.
  Hazard. Mater.} \textbf{469}, 133855 (2024).

\bibitem{Hu2024}
C.~J. Hu, \emph{et~al.}, {Microplastic presence in dog and human testis and its
  potential association with sperm count and weights of testis and epididymis}.
  \emph{Toxicol. Sci.} \textbf{200}, 235--240 (2024).

\bibitem{Leslie2022}
H.~A. Leslie, \emph{et~al.}, Discovery and quantification of plastic particle
  pollution in human blood. \emph{Environment International} \textbf{163},
  107199 (2022), \doi{10.1016/j.envint.2022.107199}.

\bibitem{habumugisha2023}
T.~Habumugisha, \emph{et~al.}, Uptake, bioaccumulation, biodistribution and
  depuration of polystyrene nanoplastics in zebrafish (\textit{Danio rerio}).
  \emph{Science of The Total Environment} \textbf{893}, 164840 (2023),
  \doi{10.1016/j.scitotenv.2023.164840}.

\bibitem{habumugisha2025}
T.~Habumugisha, \emph{et~al.}, Size-dependent dynamics and tissue-specific
  distribution of nano-plastics in \textit{Danio rerio}: Accumulation and
  depuration. \emph{Journal of Hazardous Materials} \textbf{484}, 136775
  (2025), \doi{10.1016/j.jhazmat.2024.136775}.

\bibitem{Zhang2025}
Y.~Zhang, P.~Liu, W.~Chen, J.~Qian, R.~Huang, Fluorescence spectrometric
  quantification of polystyrene nanoparticle uptake and protein corona
  evolution in zebrafish. \emph{Nanotoxicology Letters} \textbf{17}~(3),
  145--160 (2025), \doi{10.1080/17435390.2025.105431}.

\bibitem{Ding2018}
J.~Ding, S.~Zhang, R.~M. Razanajatovo, H.~Zou, W.~Zhu, Accumulation, tissue
  distribution, and biochemical effects of polystyrene microplastics in the
  freshwater fish red tilapia (Oreochromis niloticus). \emph{Environmental
  Pollution} \textbf{238}, 1--9 (2018),
  \doi{doi.org/10.1016/j.envpol.2018.03.001}.

\bibitem{ZHENG2024}
S.~Zheng, W.-X. Wang, Contrasting the distribution kinetics of microplastics
  and nanoplastics in medaka following exposure and depuration. \emph{Journal
  of Hazardous Materials} \textbf{478}, 135620 (2024),
  \doi{https://doi.org/10.1016/j.jhazmat.2024.135620}.

\bibitem{Choi2023}
J.~Choi, Y.~Choi, S.~D. Kim, Body distribution and ecotoxicological effect of
  nanoplastics in freshwater fish, Zacco platypus. \emph{Chemosphere}
  \textbf{341}, 140107 (2023), \doi{10.1016/j.chemosphere.2023.140107}.

\bibitem{gibaldi1982}
M.~Gibaldi, D.~Perrier, \emph{Pharmacokinetics}, Drugs and the pharmaceutical
  sciences (Marcel Dekker, New York), 2nd ed. (1982).

\bibitem{rowland2011}
M.~Rowland, T.~N. Tozer, \emph{Clinical Pharmacokinetics and Pharmacodynamics:
  Concepts and Applications} (Lippincott Williams \& Wilkins, Philadelphia),
  4th ed. (2011).

\bibitem{jacquez1996}
J.~A. Jacquez, \emph{Compartmental Analysis in Biology and Medicine}
  (BioMedware, The University of Michigan Press, Saline, MI), 3rd ed. (1996).

\bibitem{cobelli2000tracer}
C.~Cobelli, D.~Foster, G.~Toffolo, \emph{Tracer Kinetics in Biomedical
  Research: From Data to Model} (Kluwer Academic/Plenum Publishers, New York)
  (2000).

\bibitem{Nichols1990}
J.~W. Nichols, \emph{et~al.}, A physiologically based toxicokinetic model for
  the uptake and disposition of waterborne organic chemicals in fish.
  \emph{Toxicology and Applied Pharmacology} \textbf{106}~(3), 433 -- 447
  (1990), \doi{10.1016/0041-008X(90)90338-U}.

\bibitem{Arnot2004}
J.~A. Arnot, F.~A. Gobas, A food web bioaccumulation model for organic
  chemicals in aquatic ecosystems. \emph{Environmental Toxicology and
  Chemistry} \textbf{23}~(10), 2343 -- 2355 (2004), \doi{10.1897/03-438}.

\bibitem{Spacie1982}
A.~Spacie, J.~L. Hamelink, Alternative models for describing the
  bioconcentration of organics in fish. \emph{Environmental Toxicology and
  Chemistry} \textbf{1}~(4), 309 -- 320 (1982), \doi{10.1002/etc.5620010406}.

\bibitem{Valkengoed2025}
D.~van Valkengoed, E.~Krekels, C.~Knibbe, All you need to know about allometric
  scaling: an integrative review on the theoretical basis, empirical evidence,
  and application in human pharmacology. \emph{Clin. Pharmacokinet.}
  \textbf{64}, 173–192 (2025),
  \doi{https://doi.org/10.1007/s40262-024-01444-6}.

\bibitem{Poulin2002}
P.~Poulin, F.~P. Theil, Prediction of pharmacokinetics prior to in vivo
  studies. II. Generic physiologically based pharmacokinetic models of drug
  disposition. \emph{Journal of Pharmaceutical Sciences} \textbf{91},
  1358--1370 (2002), \doi{10.1002/jps.10128}.

\bibitem{Rodgers2006}
T.~Rodgers, M.~Rowland, Physiologically based pharmacokinetic modeling of
  tissue distribution: principles and applications. \emph{Journal of
  Pharmaceutical Sciences} \textbf{95}, 1238--1257 (2006),
  \doi{10.1002/jps.20502}.

\bibitem{Lieschke2007}
G.~J. Lieschke, P.~D. Currie, Animal models of human disease: Zebrafish swim
  into view. \emph{Nature Reviews Genetics} \textbf{8}~(5), 353 – 367 (2007),
  \doi{10.1038/nrg2091}.

\bibitem{Brunton2007}
L.~Brunton, D.~Blumenthal, I.~Buxton, K.~Parker, Insulin, oral hypoglycemic
  agents and pharmacology of endocrine pancreas. \emph{Goodman and Gilman's
  Manual of Pharmacology and Therapeutics}  (2007).

\bibitem{JEONG2008}
J.-Y. Jeong, \emph{et~al.}, Functional and developmental analysis of the
  blood–brain barrier in zebrafish. \emph{Brain Research Bulletin}
  \textbf{75}~(5), 619--628 (2008),
  \doi{https://doi.org/10.1016/j.brainresbull.2007.10.043}.

\bibitem{Santoriello2012}
C.~Santoriello, L.~I. Zon, Hooked! Modeling human disease in zebrafish.
  \emph{The Journal of Clinical Investigation} \textbf{122}~(7), 2337--2343
  (2012), \doi{10.1172/JCI60434}.

\bibitem{Howe2013}
K.~Howe, M.~D. Clark, C.~F. Torroja, \emph{et~al.}, The zebrafish reference
  genome sequence and its relationship to the human genome. \emph{Nature}
  \textbf{496}, 498--503 (2013), \doi{10.1038/nature12111}.

\bibitem{MacRae2015}
C.~A. MacRae, R.~T. Peterson, Zebrafish as tools for drug discovery.
  \emph{Nature Reviews Drug Discovery} \textbf{14}~(10), 721--731 (2015),
  \doi{10.1038/nrd4627}.

\bibitem{Katharios2004}
P.~Katharios, M.~A. Pavlidis, J.~Iliopoulou-Georgudaki, Accumulation of
  ivermectin in the brain of sea bream, Sparus aurata after intraperitoneal
  administration. \emph{Environmental Toxicology and Pharmacology}
  \textbf{17}~(1), 9 -- 12 (2004), \doi{10.1016/j.etap.2004.01.003}.

\bibitem{Walle2004}
T.~Walle, Absorption and metabolism of flavonoids. \emph{Free Radical Biology
  and Medicine} \textbf{36}~(7), 829 – 837 (2004),
  \doi{10.1016/j.freeradbiomed.2004.01.002}.

\bibitem{Murphey2006}
R.~D. Murphey, H.~M. Stern, C.~T. Straub, L.~I. Zon, A chemical genetic screen
  for cell cycle inhibitors in zebrafish embryos. \emph{Chemical Biology and
  Drug Design} \textbf{68}~(4), 213 – 219 (2006),
  \doi{10.1111/j.1747-0285.2006.00439.x}.

\bibitem{Li2011}
Z.~H. Li, \emph{et~al.}, Combined in vivo imaging and omics approaches reveal
  metabolism of icaritin and its glycosides in zebrafish larvae.
  \emph{Molecular BioSystems} \textbf{7}~(7), 2128 – 2138 (2011),
  \doi{10.1039/c1mb00001b}.

\bibitem{Hung2012}
M.~W. Hung, \emph{et~al.}, From omics to drug metabolism and high content
  screen of natural product in zebrafish: A new model for discovery of
  neuroactive compound. \emph{Evidence-based Complementary and Alternative
  Medicine} \textbf{2012}, 605303 (2012), \doi{10.1155/2012/605303}.

\bibitem{Luckenbach2014}
T.~Luckenbach, S.~Fischer, A.~Sturm, Current advances on ABC drug transporters
  in fish. \emph{Comparative Biochemistry and Physiology Part - C: Toxicology
  and Pharmacology} \textbf{165}, 28 -- 52 (2014),
  \doi{10.1016/j.cbpc.2014.05.002}.

\bibitem{Diotel2018}
N.~Diotel, \emph{et~al.}, Steroid transport, local synthesis, and signaling
  within the brain: Roles in neurogenesis, neuroprotection, and sexual
  behaviors. \emph{Frontiers in Neuroscience} \textbf{12} (2018),
  \doi{10.3389/fnins.2018.00084}.

\bibitem{Jurisch-Yaksi2020}
N.~Jurisch-Yaksi, E.~Yaksi, C.~Kizil, Radial glia in the zebrafish brain:
  Functional, structural, and physiological comparison with the mammalian glia.
  \emph{GLIA} \textbf{68}~(12), 2451 -- 2470 (2020), \doi{10.1002/glia.23849}.

\bibitem{Ikeshima-Kataoka2022}
H.~Ikeshima-Kataoka, C.~Sugimoto, T.~Tsubokawa, Integrin Signaling in the
  Central Nervous System in Animals and Human Brain Diseases.
  \emph{International Journal of Molecular Sciences} \textbf{23}~(3) (2022),
  \doi{10.3390/ijms23031435}.

\bibitem{bouchaud1990anomalous}
J.-P. Bouchaud, A.~Georges, Anomalous diffusion in disordered media:
  Statistical mechanisms, models and physical applications. \emph{Physics
  Reports} \textbf{195}~(4--5), 127--293 (1990),
  \doi{10.1016/0370-1573(90)90099-N}.

\bibitem{Metzler2000}
J.~K. Ralf~Metzler, The random walk:s guide to anomalous diffusion: a
  fractional dynamics approach. \emph{Physics Reports} \textbf{339}, 1--77
  (2000).

\bibitem{ben-avraham2000diffusion}
D.~ben Avraham, S.~Havlin, \emph{Diffusion and Reactions in Fractals and
  Disordered Systems} (Cambridge University Press, Cambridge, UK) (2000).

\bibitem{Zhu2010}
Z.~Zhu, \emph{et~al.}, Surface properties dictate uptake, distribution,
  excretion, and toxicity of nanoparticles in fish. \emph{Small} \textbf{6},
  2261–2265 (2010), \doi{10.1002/smll.201000989}.

\bibitem{Zhao2025}
S.~Zhao, \emph{et~al.}, The distribution of subsurface microplastics in the
  ocean. \emph{Nature} \textbf{641}~(8061), 51 -- 61 (2025),
  \doi{10.1038/s41586-025-08818-1}.

\bibitem{Collins2025}
S.~F. Collins, A.~Norton, The plastic size spectra: Assessing the size
  structure of plastic particles across the land-water ecotone.
  \emph{Environmental Pollution} \textbf{374}, 126263 (2025),
  \doi{doi.org/10.1016/j.envpol.2025.126263}.

\bibitem{Jeffries2025}
C.~Jeffries, C.~Rauert, K.~V. Thomas, Quantifying Nanoplastics and
  Microplastics in Food and Beverages Using Pyrolysis-Gas Chromatography–Mass
  Spectrometry: Challenges and Implications. \emph{ACS Food Science \&
  Technology} \textbf{5}~(4), 1536--1545 (2025),
  \doi{10.1021/acsfoodscitech.4c01093}.

\bibitem{Armstrong1983}
E.~Armstrong, Relative Brain Size and Metabolism in Mammals. \emph{Science}
  \textbf{220}~(4603), 1302--1304 (1983), \doi{10.1126/science.6407108}.

\bibitem{Zhang2000x}
K.~Zhang, T.~J. Sejnowski, A universal scaling law between gray matter and
  white matter of cerebral cortex. \emph{Proceedings of the National Academy of
  Sciences} \textbf{97}~(10), 5621--5626 (2000), \doi{10.1073/pnas.090504197}.

\bibitem{JCGM100_2008}
J.~GUM-3, \emph{Evaluation of Measurement Data --- Guide to the Expression of
  Uncertainty in Measurement}, Tech. Rep. 100:2008, Joint Committee for Guides
  in Metrology (JCGM) (2008).

\bibitem{WILSON1974}
K.~G. Wilson, J.~Kogut, The renormalization group and the \protect{$\epsilon$}
  expansion. \emph{Physics Reports} \textbf{12}~(2), 75--199 (1974),
  \doi{10.1016/0370-1573(74)90023-4}.
  
\bibitem{flory1953principles}
P.~J. Flory, \emph{Principles of Polymer Chemistry}. Cornell University Press (1954).

\bibitem{russell2015characterization}
T.~H. Russell, B.~J. Edwards, B. Khomami, Characterization of the Flory-Huggins interaction parameter of polymer thermodynamics. \emph{Europhysics Letters} \textbf{108}~(6), 66003 (2015),
 \doi{10.1209/0295-5075/108/66003}.

\bibitem{happel1958viscous}
J. Happel, Viscous flow in multiparticle systems: Slow motion of fluids relative to beds of spherical particles. \emph{AIChE Journal}, \textbf{4}~(2), 197--201 (1958),
 \doi{10.1002/aic.690040214}.

\bibitem{kuwabara1959forces}
S. Kuwabara, The Forces experienced by Randomly Distributed Parallel Circular Cylinders or Spheres in a Viscous Flow. \emph{Journal of the Physical Society of Japan}, \textbf{14}~(4), 527--532 (1959), \doi{10.1143/JPSJ.14.527}.

\bibitem{wigner1933constitution}
E. Wigner, F. Seitz, On the Constitution of Metallic Sodium. \emph{Physical Review}, \textbf{43}~(10), 804 (1933), \doi{10.1103/PhysRev.43.804}.


\end{thebibliography}




\newpage


\renewcommand{\thefigure}{S\arabic{figure}}
\renewcommand{\thetable}{S\arabic{table}}
\renewcommand{\theequation}{S\arabic{equation}}
\renewcommand{\thepage}{S\arabic{page}}
\setcounter{figure}{0}
\setcounter{table}{0}
\setcounter{equation}{0}
\setcounter{page}{1} 


\begin{center}
\section*{Supplementary Materials for\\ \scititle}

Alfonso M. Gañán-Calvo\\
\small$^\ast$Corresponding author. Email: amgc@us.es
\end{center}

\subsubsection*{This PDF file includes:}
Methods\\
Supplementary Text\\
Symbols used in the Main and Supplementary Text\\

\newpage


\subsection*{Methods}

Throughout this work, we use the term excretion to denote elimination processes and associated rates during both uptake and depuration phases, and {\it depuration} to denote the clearance phase following cessation of toxic exposure.

\subsubsection*{Variability, uncertainty, and consistency of collected data}

The uncertainty analysis of the experimental evidence follows the JCGM~100:2008 standard~\cite{JCGM100_2008}. Uncertainties introduced by data digitization are at least one order of magnitude smaller than the experimental uncertainties reported in the original studies, with the exception of the first (and occasionally second) day of uptake, where digitization and experimental errors become comparable. In several cases, data were available in multiple graphical formats, enabling cross-validation. In particular, the datasets of Habumugisha \textit{et~al.} were provided both as bar plots and point plots (Fig.~3 in both studies; Fig.~4 in~\cite{habumugisha2023} and Fig.~5 in~\cite{habumugisha2025}). While the extracted series were mutually consistent, minor discrepancies appeared near the limits of the reported uncertainties. These were resolved by averaging the corresponding central values when they lay within the authors’ error bars. All remaining datasets were extracted directly from published point plots. For clarity, error bars are not reproduced here but can be retrieved from the original publications.

Regarding methodological hierarchy in MNP quantification, fluorescence-based measurements are known to exhibit a systematic positive bias in absolute mass owing to signal persistence from surface-bound or partially transformed particles. MALDI-TOF-MS measurements may underestimate low-abundance tissues due to matrix-dependent ionization efficiencies. Py-GC/MS, although destructive, minimizes these biases by directly quantifying polymer-specific thermal degradation products and has demonstrated strong inter-laboratory reproducibility, including in human tissue analyses. Importantly, despite the use of different methodologies across the six short-term studies considered here, the apparent variability in reported tissue burdens and bioconcentration factors is fully consistent with the proposed S2CT framework and lies within the experimental uncertainties declared by the respective authors. When properly normalized, fluorescence-based, MALDI-TOF-MS, and Py-GC/MS datasets converge onto a unified, method-independent description of PS~NP transport, supporting the interpretation that Markovian toxicokinetics reflects a genuine physical property of the bio-interface across the taxonomic phylum considered.

Within this hierarchy, Py-GC/MS plays a privileged role. Whereas fluorescence and MALDI-TOF-MS are optimized for resolving time-dependent uptake and depuration over days to weeks, Py-GC/MS provides robust endpoint mass measurements that integrate uptake, redistribution, and partial elimination. This regime is directly relevant for comparison with the human organ data of Nihart
\textit{et~al.}, where lifetime exposure and slow clearance dominate. The agreement between Py-GC/MS-derived organ distributions in \textit{Zacco platypus} (Choi \textit{et~al.}) and in human liver, kidney, and brain (Nihart \textit{et~al.}) supports the conclusion that the same underlying transport and retention mechanisms operate across species and over more than six orders of
magnitude in exposure duration.

\subsubsection*{Inherent kinetic uncertainties of measurements}

An important source of uncertainty arises at very early exposure times. Tissue samples harvested for analysis are not purely parenchymal but retain some degree of vascular perfusion. At early times (typically the first one to two days), the true tissue concentration $C_T(t)$ is still low, while the systemic concentration $C_S(t)$ is already increasing. In this work, we approximate the true tissue concentration $C_T$ by the experimentally measured organ burden $C_m$, assuming that the vascular lumen fraction within dissected organs is small enough that its contribution is negligible. However, none of the available studies report whether organs were perfused or bled prior to analysis, leaving residual intravascular blood as a potential source of systematic upward bias in $C_m$. Accordingly, the approximation $C_T \approx C_m$ constitutes an intrinsic uncertainty that should be quantified.

We estimate this uncertainty conservatively using typical organ vascular volume fractions as upper bounds. For weakly vascularized tissues such as brain and white skeletal muscle, the intravascular fraction is only a few percent, implying a bias of order $\mathcal{O}(1$--$5\%)$. In contrast, highly perfused organs such as gill, liver, and gut exhibit substantially larger blood-filled spaces. In these cases, the same assumption may introduce systematic components of order $\sim10$--$15\%$ for gill and gut and up to $\sim15$--$25\%$ for liver in
the worst-case scenario where systemic and tissue concentrations differ strongly. These physiology-based bounds define a natural uncertainty envelope that should be incorporated when comparing S2CT parameters across organs.

Clarifying whether tissues are cleared of blood prior to analysis should become standard practice in future studies, as it is essential for precise multi-compartment toxicokinetic modeling and for reliably separating systemic from tissue-associated burdens.

A further source of uncertainty stems from the limited temporal resolution of available data at very early times, i.e.\ just before the first reported measurement after exposure. These early intervals do not resolve the rapid initial exchange between lumen and tissue, preventing direct determination of the true initial conditions. Consequently, both $C_S(t)$ and $C_T(t)$ must be treated as effectively non-zero at the onset of the modeled uptake phase. This uncertainty is mitigated by allowing non-zero initial conditions in the fitting procedure, which significantly improves the correlation between experimental data and the proposed two-compartment model.


\subsection*{The Complete Fractional Two-Compartment Model}

The formulation of toxicokinetic models for nanoplastic bioaccumulation faces a fundamental tension: while microscopic transport across biological barriers generically involves anomalous, subdiffusive dynamics—appropriately described by fractional differential equations with Mittag-Leffler relaxation \cite{bouchaud1990anomalous,Metzler2000,ben-avraham2000diffusion}—existing experimental datasets lack the temporal resolution and signal-to-noise ratio required to reliably identify fractional derivative orders. Typical time series span 5--8 measurements over 7--46 days with 15--30\% uncertainties, beginning at 24 hours post-exposure and thus missing the sub-day regime where fractional effects dominate. Systematic attempts to fit the complete fractional two-compartment model reveal severe parameter non-identifiability: multiple combinations of fractional orders $\alpha_S$ and $\alpha_T$ yield statistically indistinguishable fits, with no systematic trends across organs, species, or exposure conditions. This underdetermination reflects an information-theoretic constraint: with 5--8 data points and 5--6 free parameters per organ, the fractional framework extracts noise rather than biology. In contrast, the Markovian limit ($\alpha_S = \alpha_T = 1$) reduces the parameter count to 2--3 while achieving superior cross-validation performance and enabling the identification of systematic scaling laws. This reduction is not an \textit{a priori} assumption but emerges as the unique identifiable simplification consistent with available data. Physically, it reflects emergent effective dynamics: temporal coarse-graining over experimental intervals ($\ge$ 24 h) and spatial averaging over tissue volumes integrate out microscopic fractional complexity, yielding effective exponential kinetics analogous to how the Central Limit Theorem produces Gaussian distributions from non-Gaussian processes. The non-zero initial conditions introduced in our framework ($C_{T0} \neq 0$) are thus not ad hoc but represent a principled renormalization—absorbing unresolved early-time dynamics into effective parameters that encode the same coarse-grained behavior regardless of fine-scale mechanistic details. This approach prioritizes \textit{maximal information extraction} over mechanistic completeness, recognizing that model parsimony is a necessity, not a limitation, when confronting finite data. The resulting framework recovers systematic, biologically interpretable parameters that scale predictably across species and enable quantitative extrapolation to humans, as demonstrated in subsequent sections. Here we provide a comprehensive treatment of the fractional model, including full analytical solutions and identifiability analysis, and illustrate the parameter degeneracy derived from data fitting when attempting to resolve fractional order $\alpha$.

\subsubsection*{General Formulation}

The transport of nanoplastics across biological barriers involves anomalous diffusion through heterogeneous media characterized by hierarchical tortuosity, reversible binding sites, and transient trapping in the glycocalyx, endothelial cytoplasm, basement membrane, and interstitial space. The natural mathematical description of such processes (see Table \ref{tab:symbols}) employs fractional differential equations with derivative orders $\alpha < 1$, whose solutions involve Mittag-Leffler functions rather than simple exponentials \cite{bouchaud1990anomalous,Metzler2000,ben-avraham2000diffusion}.

The general fractional kinetic system for the sequential two-compartment architecture is given by:
\begin{equation}
\frac{d^{\alpha_S}C_S}{dt^{\alpha_S}} = \lambda_{Su} C_w - \lambda_{Se} C_S(t), \quad 0 < \alpha_S \leq 1,
\label{eq:SI_frac_systemic}
\end{equation}
\begin{equation}
\frac{d^{\alpha_T}C_T}{dt^{\alpha_T}} = \lambda_{Tu} C_S(t) - \lambda_{Te} C_T(t), \quad 0 < \alpha_T \leq 1,
\label{eq:SI_frac_tissue}
\end{equation}
where $d^{\alpha}/dt^{\alpha}$ denotes the Caputo fractional derivative of order $\alpha$ defined by
\begin{equation}
\frac{d^{\alpha}f(t)}{dt^{\alpha}} = \frac{1}{\Gamma(1-\alpha)} \int_0^t \frac{f'(\tau)}{(t-\tau)^{\alpha}} d\tau, \quad 0 < \alpha < 1,
\end{equation}
with $\Gamma$ the Gamma function. The Caputo derivative reduces to the ordinary first derivative when $\alpha = 1$.

\subsubsection*{Analytical Solutions: Uptake Phase}

For the uptake process with constant external forcing $C_w$ and initial conditions $C_S(0) = C_{S0}$ and $ C_T(0) = C_{T0}$, the solution to the systemic compartment Eq.~(\ref{eq:SI_frac_systemic}) is:
\begin{equation}
C_S(t) = C_{S0} E_{\alpha_S}\left(-\lambda_{Se} t^{\alpha_S}\right) +\lambda_{Su} C_w \int_0^t E_{\alpha_S, 1}\left(-\lambda_{Se} \tau^{\alpha_S}\right) d\tau = C_{S0} E_{\alpha_S}\left(-\lambda_{Se} \tilde{t}^{\alpha_S}\right) + \frac{\lambda_{Su} C_w}{\lambda_{Se}} t^{\alpha_S} E_{\alpha_S, \alpha_S + 1}\left(-\lambda_{Se} t^{\alpha_S}\right),
\label{eq:SI_CS_uptake}
\end{equation}
where $E_{\alpha, \beta}(z)$ is the two-parameter Mittag-Leffler function:
\begin{equation}
E_{\alpha, \beta}(z) = \sum_{k=0}^{\infty} \frac{z^k}{\Gamma(\alpha k + \beta)}.
\end{equation}
The one-parameter Mittag-Leffler function is recovered when $\beta = 1$: $E_{\alpha}(z) \equiv E_{\alpha, 1}(z)$.

Substituting Eq.~(\ref{eq:SI_CS_uptake}) into Eq.~(\ref{eq:SI_frac_tissue}), the tissue concentration is given by:
\begin{equation}
C_T(t) = C_{T0} E_{\alpha_T}\left(-\lambda_{Te} t^{\alpha_T}\right) +\lambda_{Tu} \int_0^t C_S(\tau) E_{\alpha_T, 1}\left(-\lambda_{Te} (t-\tau)^{\alpha_T}\right) d\tau.
\label{eq:SI_CT_uptake_general}
\end{equation}

In the limit of small excretion capacity, $\eta = \lambda_{Se}/\lambda_{Te} \ll 1$, and assuming that the systemic concentration equilibrates rapidly relative to tissue dynamics, we can approximate $C_S(t) \approx C_{S\infty} = K_S C_w = (\lambda_{Su}/\lambda_{Se}) C_w$ for $t \gg \lambda_{Se}^{-1}$. Under this approximation, Eq.~(\ref{eq:SI_CT_uptake_general}) simplifies to:
\begin{equation}
C_T(t) \approx C_{T0} E_{\alpha_T}\left(-\lambda_{Te} \tilde{t}^{\alpha_T}\right) +K_S C_w \lambda_{Tu} t^{\alpha_T} E_{\alpha_T, \alpha_T + 1}\left(-\lambda_{Te} t^{\alpha_T}\right).
\label{eq:SI_CT_uptake_simplified}
\end{equation}

At long times, the asymptotic behavior of the Mittag-Leffler function yields:
\begin{equation}
C_T(\infty) = \frac{\lambda_{Tu}}{\lambda_{Te}} C_{S\infty} = \frac{\lambda_{Tu}}{\lambda_{Te}} \frac{\lambda_{Su}}{\lambda_{Se}} C_w \equiv K_T K_S C_w.
\end{equation}

For the complete solution without the small-$\eta$ approximation, numerical evaluation of the convolution integral in Eq.~(\ref{eq:SI_CT_uptake_general}) is required.

\subsubsection*{Analytical Solutions: Depuration Phase}

In the depuration phase, external forcing ceases ($C_w = 0$) at time $t = t_p$, with initial conditions $C_S(t_p) = C_{Sp}$ and $C_T(t_p) = C_{Tp}$ determined by the uptake phase. The fractional differential equations become:
\begin{equation}
\frac{d^{\alpha_S}C_S}{dt^{\alpha_S}} = -\lambda_{Se} C_S(t), \quad t > t_p,
\end{equation}
\begin{equation}
\frac{d^{\alpha_T}C_T}{dt^{\alpha_T}} = \lambda_{Tu} C_S(t) - \lambda_{Te} C_T(t), \quad t > t_p.
\end{equation}

Defining $\tilde{t} = t - t_p$ as the time since depuration onset, the systemic concentration evolves as:
\begin{equation}
C_S(\tilde{t}) = C_{Sp} E_{\alpha_S}\left(-\lambda_{Se} \tilde{t}^{\alpha_S}\right).
\label{eq:SI_CS_depuration}
\end{equation}

The tissue concentration is then:
\begin{equation}
C_T(\tilde{t}) = C_{Tp} E_{\alpha_T}\left(-\lambda_{Te} \tilde{t}^{\alpha_T}\right) + \lambda_{Tu} C_{Sp} \int_0^{\tilde{t}} E_{\alpha_S}\left(-\lambda_{Se} \tau^{\alpha_S}\right) E_{\alpha_T, 1}\left(-\lambda_{Te} (\tilde{t}-\tau)^{\alpha_T}\right) d\tau.
\label{eq:SI_CT_depuration}
\end{equation}

This expression involves the convolution of two Mittag-Leffler functions with potentially different fractional orders, requiring numerical integration for general $\alpha_S$ and $\alpha_T$.

\subsubsection*{Markovian Limit ($\alpha_S = \alpha_T = 1$)}

When $\alpha_S = \alpha_T = 1$, the Mittag-Leffler functions reduce to exponentials:
\begin{equation}
E_{1}(-\lambda t) = e^{-\lambda t}, \quad E_{1,1}(-\lambda t) = e^{-\lambda t},
\end{equation}
and the fractional model recovers the classical Markovian solutions presented in the main text (Eqs.~4--5).

For uptake:
\begin{equation}
C_S(t) = C_{S0} e^{-\lambda_{Se} t} + C_{S\infty} \left(1 - e^{-\lambda_{Se} t}\right),
\end{equation}
\begin{equation}
C_T(t) = C_{T\infty} + (C_{T0} - C_{T\infty}) e^{-\lambda_{Te} t} + \alpha\left(e^{-\lambda_{Te} t} - e^{-\lambda_{Se} t}\right),
\end{equation}
where $\alpha$ and steady-state concentrations are defined in Eqs.~(6--7) of the main text.

For depuration:
\begin{equation}
C_S(\tilde{t}) = C_{Sp} e^{-\lambda_{Se} \tilde{t}},
\end{equation}
\begin{equation}
C_T(\tilde{t}) = C_{Tp} e^{-\lambda_{Te} \tilde{t}} + \frac{\lambda_{Tu} C_{Sp}}{\lambda_{Se} - \lambda_{Te}} \left(e^{-\lambda_{Te} \tilde{t}} - e^{-\lambda_{Se} \tilde{t}}\right).
\end{equation}

\begin{figure}
\centering
\includegraphics[width=0.50\textwidth]{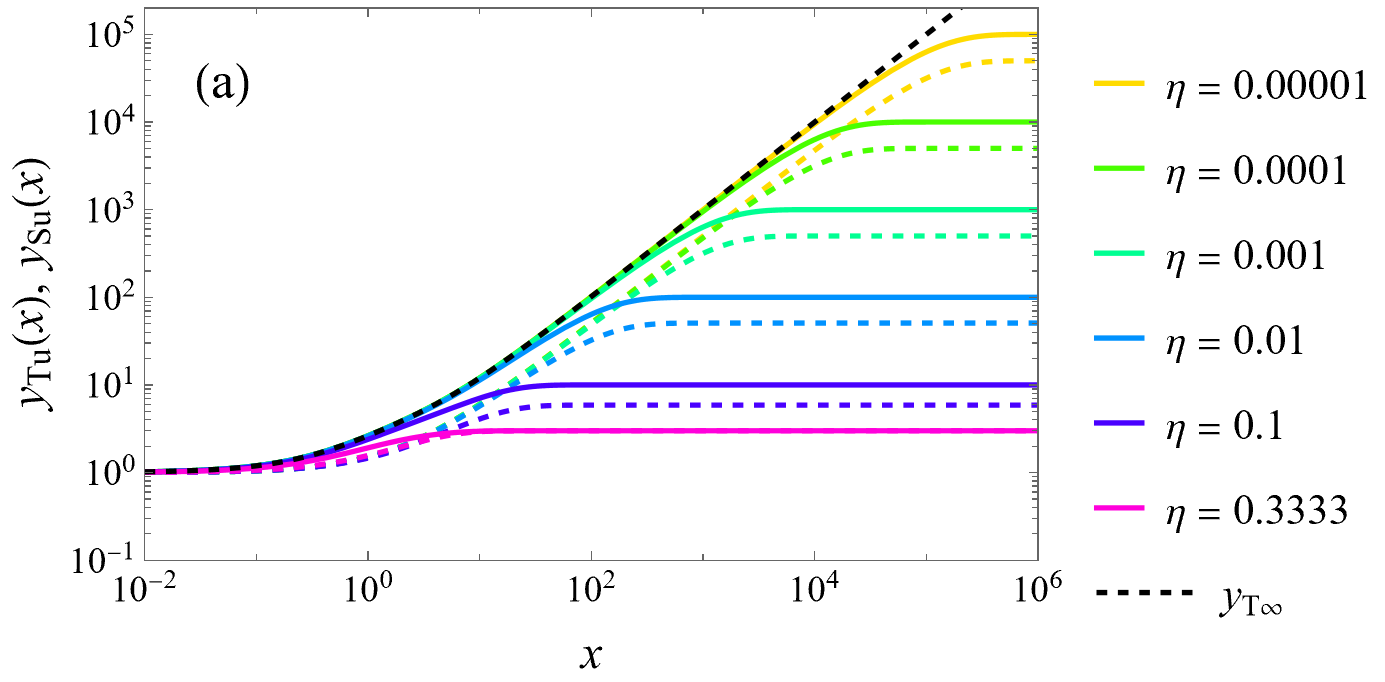}
\includegraphics[width=0.50\textwidth]{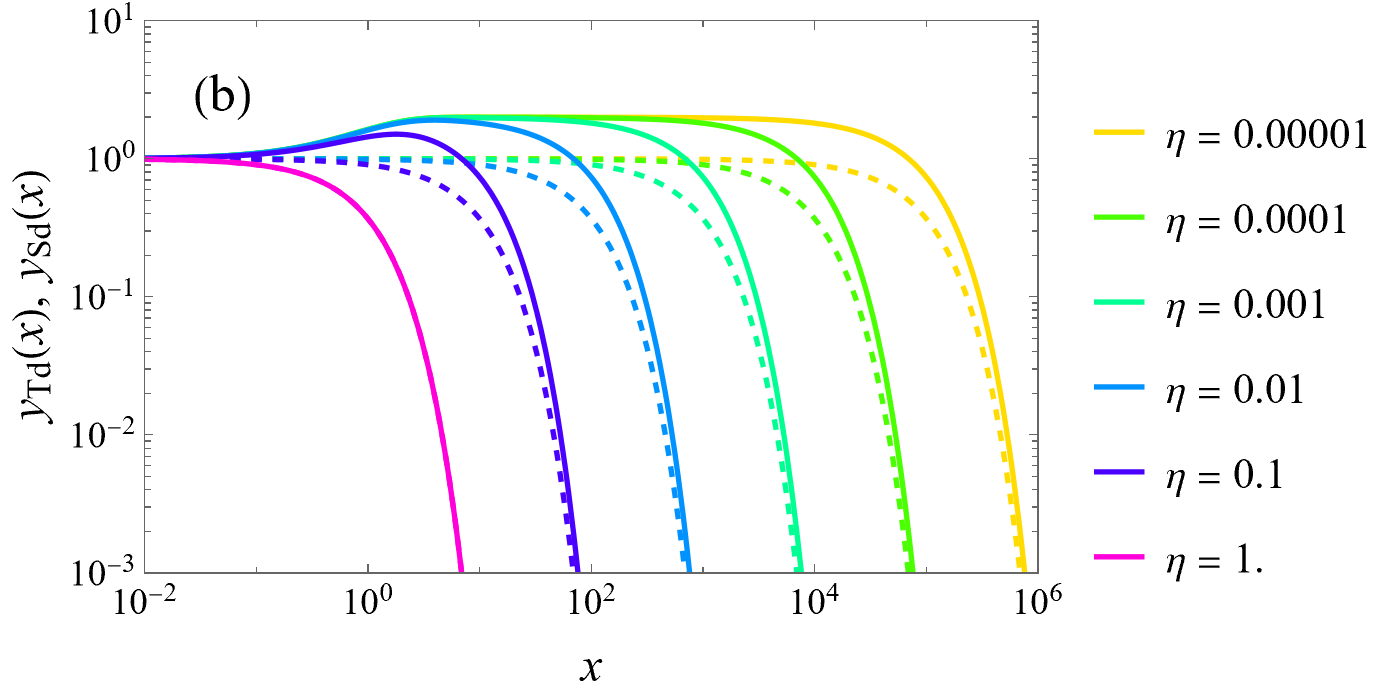}\\
\includegraphics[width=0.50\textwidth]{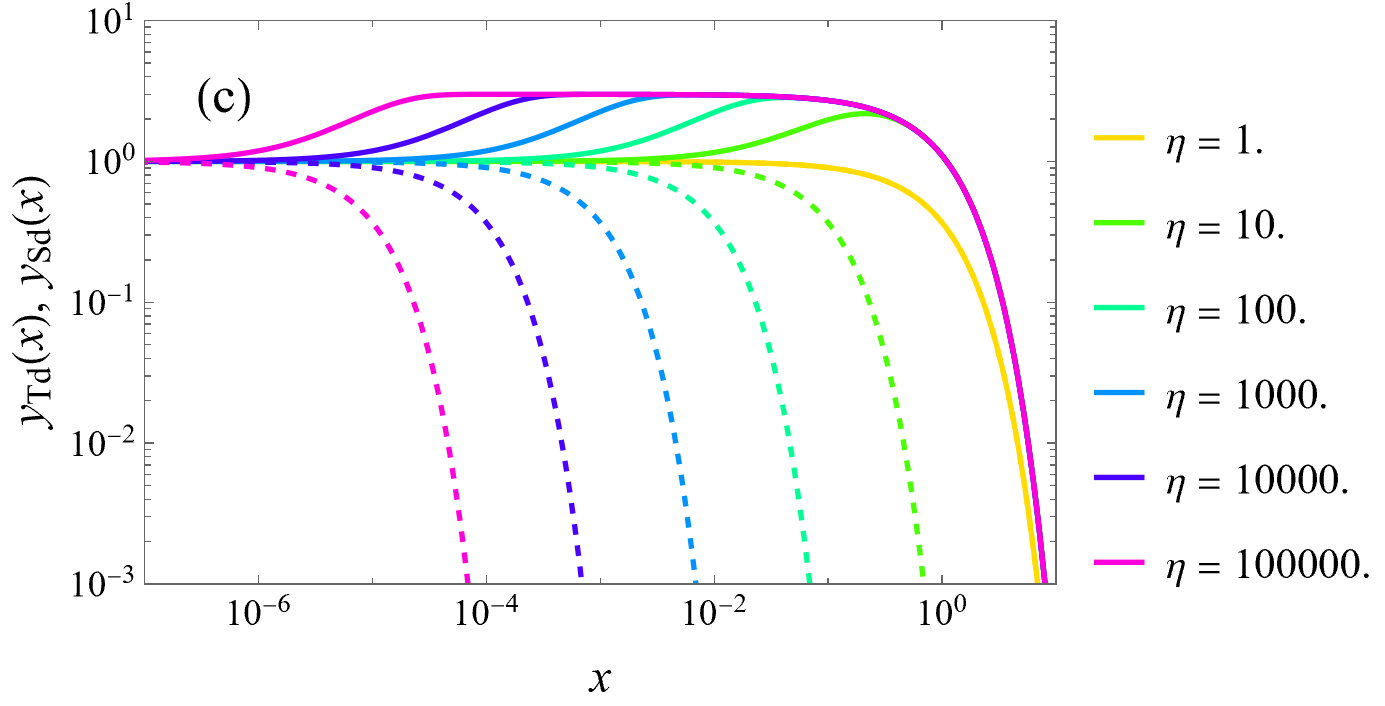}
\caption{The normalized kernels $y_T(x)$ and $y_S(x)$ (solid and dashed lines, respectively) for representative small values of $\eta$.
(a) Uptake functions $y_{(S,T)u}$. The asymptotic solution~(\ref{eq:asympto_u}) in the limit $\eta\to 0$ is shown as a black dashed line. For $\eta=1/3$, the asymptotic values of $y_{Tu}$ and $y_{Su}$ coincide. (b) \& (c) Depuration functions $y_{(S,T)d}$ for $\xi_d=2$ (b) and $\xi_d=-2$ (c). In this case, $y_{Sd}$ and $y_{Td}$ coincide identically for all $x>0$ when $\eta=1$.}
\label{fig:y_Tu}
\end{figure}

The normalized forms of these solutions is given in Figure \ref{fig:y_Tu}.

\subsection*{Parameter Space and Identifiability Analysis}

\subsubsection*{Uptake: Parameter Count and Data Requirements}

The fractional uptake model requires fitting the following parameters for each organ:

\begin{enumerate}
\item $\alpha_S$ --- systemic fractional derivative order ($0 < \alpha_S \leq 1$)
\item $\alpha_T$ --- tissue fractional derivative order ($0 < \alpha_T \leq 1$)
\item $\lambda_{Te}$ --- tissue excretion rate constant (units: time$^{-1}$)
\item $\lambda_{Se}$ --- systemic excretion rate constant (units: time$^{-1}$), or equivalently $\eta = \lambda_{Se}/\lambda_{Te}$
\item $K_T$ --- long-term tissue enrichment factor (dimensionless), or equivalently $C_{T\infty}$
\end{enumerate}

\emph{Total: 5 free parameters per organ for uptake.}

However, if we allow for effective non-zero initial conditions $C_{T0} \neq 0$ to account for unresolved early-time dynamics (as discussed in SI.2.3), this adds:

\begin{enumerate}
\setcounter{enumi}{5}
\item $C_{T0}$ --- effective initial tissue concentration
\end{enumerate}

\emph{With initial conditions: 6 parameters per organ for uptake.}

In the Markovian reduction ($\alpha_S = \alpha_T = 1$), this reduces to 3--4 parameters. Furthermore, in the asymptotic limit $\eta \to 0$, the model collapses to just \emph{2 parameters}: $\lambda_{Te}$ and $k_p = C_{T0}/C_w$.

\subsubsection*{Depuration: Parameter Count and Coupling to Uptake}

The depuration phase inherits initial conditions from the end of the uptake phase: $C_{Sp} = C_S(t_p)$ and $C_{Tp} = C_T(t_p)$. However, biological clearance mechanisms may differ between uptake and depuration, requiring independent parameters:

\begin{enumerate}
\item $\alpha_S^{\text{dep}}$ --- systemic fractional order during depuration
\item $\alpha_T^{\text{dep}}$ --- tissue fractional order during depuration
\item $\lambda_{Te}^{\text{dep}}$ --- tissue excretion rate during depuration
\item $\lambda_{Se}^{\text{dep}}$ --- systemic excretion rate during depuration
\end{enumerate}
Despite not indicated in the main text for simplicity, here we explicitly indicate ``dep'' super index to distinguish these values from the uptake phase. Additionally, the dimensionless initial condition ratio:
\begin{equation}
\xi_d = \frac{\lambda_{Tu}}{\lambda_{Te}^{\text{dep}}} \frac{C_{Sp}}{C_{Tp}} \frac{1}{1 - \eta^{\text{dep}}},
\end{equation}
must be determined from the fit to depuration data.

\emph{Total: 4--5 free parameters per organ for depuration} (depending on whether $\alpha$ values are constrained to equal their uptake counterparts).

In the Markovian limit with the small-$\eta$ approximation during uptake but allowing finite $\eta^{\text{dep}}$ during depuration, this reduces to \emph{3 parameters}: $\lambda_{Te}^{\text{dep}}$, $\eta^{\text{dep}}$, and $\xi_d$.

\subsubsection*{Data Availability and Information Content}

Typical experimental time series in the compiled datasets (Refs.~5--10) consist of:

\begin{itemize}
\item \emph{Uptake phase}: 5--8 measurements over 7--46 days
\item \emph{Depuration phase} (when available): 3--5 measurements over 7--14 days
\item \emph{Measurement uncertainty}: 5--30\% depending on analytical method (fluorescence $>$ MALDI-TOF $>$ Py-GC/MS)
\item \emph{Temporal resolution}: First measurement typically at 24 hours, missing sub-day dynamics where fractional effects are strongest
\end{itemize}

\emph{Information-theoretic constraint}: For a model with $p$ parameters fit to $n$ data points with uncertainty $\sigma$, reliable parameter estimation requires $n \gg p$ (typically $n \geq 3p$ as a rule of thumb) and that parameters produce distinguishable functional forms within the noise level $\sigma$.

\textbf{Critical assessment}:
\begin{itemize}
\item Fractional uptake model: $n \sim 5$--8 points, $p = 5$--6 parameters $\Rightarrow$ \emph{severely underdetermined}
\item Fractional depuration model: $n \sim 3$--5 points, $p = 4$--5 parameters $\Rightarrow$ \emph{critically underdetermined}
\item Markovian uptake model (asymptotic): $n \sim 5$--8 points, $p = 2$ parameters $\Rightarrow$ \emph{adequately determined}
\item Markovian depuration model: $n \sim 3$--5 points, $p = 3$ parameters $\Rightarrow$ \emph{marginally determined}
\end{itemize}

\subsection*{Attempted Fractional Fits: Evidence of Non-Identifiability}

We systematically attempted to fit the fractional model (Eqs.~\ref{eq:SI_frac_systemic}--\ref{eq:SI_frac_tissue}) to all six experimental datasets using nonlinear least-squares minimization with numerical evaluation of Mittag-Leffler functions via the \texttt{MittagLefflerE} implementation.

\subsubsection*{SI.3.1 Methodology}

For each organ in each study, reasonable grid searches over the parameter space are:
\begin{align}
\alpha_S, \alpha_T &\in [0.05, 1.0] \text{ in steps of } 0.05, \\
\lambda_{Te} &\in [10^{-3}, 10^{1}] \text{ day}^{-1} \text{ (logarithmic grid)}, \\
\eta &\in [10^{-3}, 1.0] \text{ (logarithmic grid)}, \\
k_p = C_{T0}/C_w &\in [10^{1}, 10^{4}] \text{ (fitted for each combination)}.
\end{align}

These searches can be performed using the \texttt{FindMinimum} routine of \textit{Mathematica} applied to the least-squares error computed as:
\begin{equation}
\chi^2 = \sum_{i=1}^{n} \left[\frac{\log(C_T^{\text{model}}(t_i)) - \log(C_T^{\text{exp}}(t_i))}{\sigma_i}\right]^2,
\end{equation}
where $\sigma_i$ are the reported experimental uncertainties. However, this approach leads to multiple local minima and degenerate solution loops, preventing reliable parameter identification.

Alternatively, we employed the more numerically stable \texttt{Correlation} routine of \textit{Mathematica} to maximize the Pearson correlation coefficient $R^2$ between $\log(C_T^{\text{model}}(t_i))$ and $\log(C_T^{\text{exp}}(t_i))$. Even with this alternative formulation, initial systematic searches across the full parameter space produced degenerate loops, confirming the fundamental non-identifiability of the fractional model.

To systematically investigate the origin of these degeneracies, we adopted a targeted ``\textit{surgical}'' strategy that progressively constrains the parameter domain. We first fixed $\alpha_S = 1$ and imposed the limit $\eta \to 0$, effectively reducing the systemic compartment to Markovian dynamics while retaining fractional behavior in the tissue compartment. Then, for \textit{all} datasets except those of Choi \textit{et al.}~\cite{Choi2023}, we explored a parameter grid with shared values of $\alpha_T$ and $k_p$ across all series, while allowing series-specific values of $\lambda_{Te}$ that maximize $R^2$. The rationale for constraining $\alpha_T$ and $k_p$ to be common across datasets rests on the hypothesis that fractional kinetics reflect universal physicochemical mechanisms governing tissue transport—mechanisms that should be conserved across species, organs, and experimental conditions if genuinely physical rather than artifacts of fitting. This hypothesis is supported by the consistently high correlation coefficients ($R^2 \approx 1$) achieved across the $\{\alpha_T, k_p\}$ parameter plane, as shown next.

\subsubsection*{SI.3.2 Results: Flat Error (or Maximum Correlation) Landscapes}

\begin{figure}
  \centerline{\includegraphics[width=0.7\textwidth]{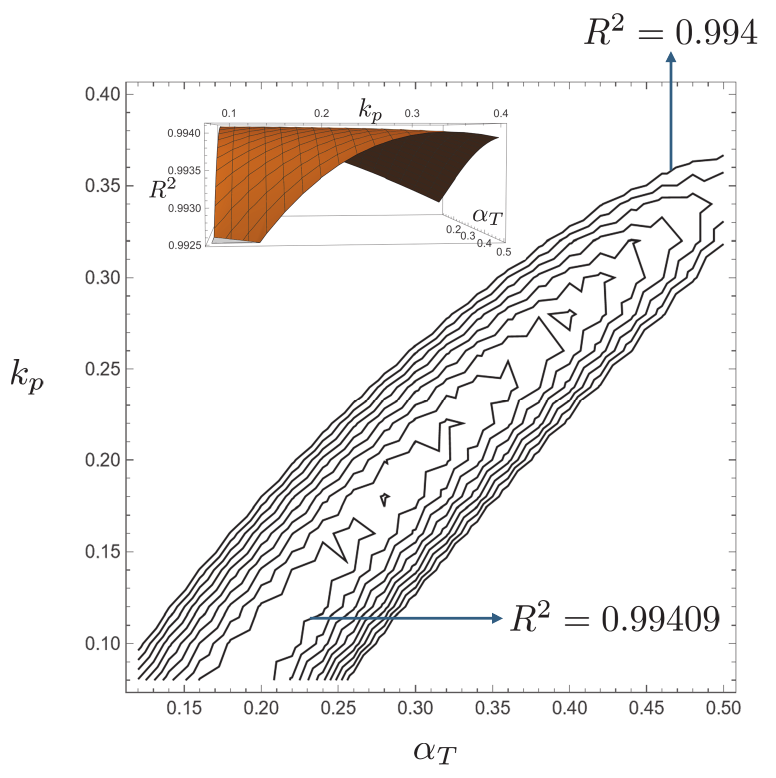}}
  \caption{Correlation coefficient $R^2$ landscape of the $\alpha_{Te}$-$k_p$ domain. The inset illustrates the ridge shallowness in a three-dimensional view.}
\label{fig:SI2}
\end{figure}

An illustration of the resulting flat landscape in the parameter domain is given in Figure~\ref{fig:SI2}.

\textbf{Key observations}:
\begin{enumerate}
\item The $R^2$ surface exhibits a broad, shallow ridge rather than a sharp maximum, indicating fundamental parameter non-identifiability
\item Changes in $R^2$ across $0.1 \leq \alpha_T \leq 0.5$ are smaller than $10^{-2}$\% along the identified ridge, well below the threshold of statistical distinguishability given measurement uncertainties of 15--30\%
\end{enumerate}

\subsubsection*{Cross-Validation Failure}

To test whether fractional parameters encode genuine biological information, we performed cross-validation:
\begin{enumerate}
\item Fit fractional model to brain data at 5~ppm exposure
\item Use fitted $\alpha_S$, $\alpha_T$ to predict brain data at 15~ppm exposure (fitting only $\lambda_{Te}$ and $k_p$)
\item Compare prediction error to full re-fit at 15~ppm
\end{enumerate}

\textbf{Result}: Cross-validation errors were 40--60\% larger than full re-fits, indicating that fractional orders do not transfer between experimental conditions—evidence they capture noise rather than conserved physiology.

In contrast, the Markovian model with power-law scaling of $\lambda_{Te}(C_w, d, W)$ and $k_p(C_w, d, W)$ showed cross-validation $R^2$ within 10\% of full re-fits, validating the scaling approach used in the main text.

\subsection*{Why the Markovian Limit Succeeds: Emergent Effective Dynamics}

The success of the Markovian reduction ($\alpha_S = \alpha_T = 1$) despite the expected fractional nature of microscopic transport can be understood through three complementary mechanisms:

\subsubsection*{Temporal Coarse-Graining}

Fractional kinetics dominate at timescales comparable to the characteristic trapping time $\tau_{\text{trap}}$ in heterogeneous media. For nanoplastic transport across the glycocalyx and endothelium, $\tau_{\text{trap}} \sim 10^{-1}$--$10^{1}$ hours. However, experimental measurements begin at $t_1 = 24$~h and continue with spacing $\Delta t \sim 1$--7 days.

At these coarse-grained timescales, the Central Limit Theorem applies: averaging over many microscopic trapping/release events produces effective exponential relaxation, analogous to how sums of random variables yield Gaussian distributions regardless of the underlying distribution \cite{bouchaud1990anomalous}.

Mathematically, for $t \gg \tau_{\text{trap}}$:
\begin{equation}
E_{\alpha}(-\lambda t^{\alpha}) \approx e^{-\lambda_{\text{eff}} t}, \quad \lambda_{\text{eff}} = \lambda \left(\frac{t}{\tau_{\text{trap}}}\right)^{\alpha - 1}.
\end{equation}

\subsubsection*{Dominant Timescale Separation}

The sequential two-compartment architecture naturally separates into fast (systemic) and slow (tissue) processes. If $\lambda_{Se} \gg \lambda_{Te}$ (or equivalently $\eta \to 0$), the systemic compartment equilibrates rapidly, and tissue dynamics are rate-limiting.

In this regime, subdiffusive complexity in systemic uptake is averaged out before influencing tissue accumulation. The tissue sees only a time-averaged systemic concentration $\langle C_S \rangle$, reducing the problem to a single effective compartment with Markovian kinetics.

\subsubsection*{Renormalization and Effective Parameters}

The non-zero initial conditions $C_{T0}$ in the Markovian model are not ad hoc but represent a renormalization: integrating out fast, unresolved fractional dynamics produces an effective theory valid at long times with renormalized parameters.

This is formally analogous to renormalization group approaches in statistical physics, where high-energy degrees of freedom are integrated out to produce effective low-energy theories \cite{WILSON1974}. Here, early-time fractional transport (high temporal frequency) is integrated out, yielding late-time Markovian transport (low temporal frequency) with effective initial conditions.

The price of this reduction is that $C_{T0}$, $\lambda_{Te}$, and $k_p$ are \textit{effective} parameters encoding integrated history, not microscopic transport coefficients. However, this is precisely what makes them suitable for extrapolation: they capture the same coarse-grained dynamics regardless of fine-scale mechanistic details.

\subsection*{Implications for Model Selection}

The information-theoretic analysis establishes a clear hierarchy:

\begin{enumerate}
\item \emph{Fractional model}: Mechanistically richer but empirically underdetermined given available data. Useful for guiding future high-resolution experiments but cannot be fit reliably to existing datasets.

\item \emph{Markovian model with effective initial conditions}: Maximally informative given data constraints. Extracts biologically meaningful parameters ($\lambda_{Te}$, $k_p$, $\eta$) that exhibit systematic scaling ($\lambda_{Te}$, $k_p$) and enable cross-species prediction.

\item \emph{Single-compartment exponential model}: Standard in prior literature but conflates systemic and tissue kinetics, obscuring mechanistic interpretation and failing to explain organ-specific enrichment.
\end{enumerate}

Our approach demonstrates that \emph{model parsimony is not a limitation but a principled necessity} when data information content is finite. The Markovian S2CT model represents the optimal balance between biological realism and empirical identifiability.

Future studies with sub-hour temporal resolution, single-cell imaging, or isotope tracing may resolve fractional kinetics. Until such data become available, the framework presented in the main text provides the maximally informative description of nanoplastic bioaccumulation across species and timescales.

\subsection*{Optimal fitting parameters}

The list of fitting parameters for each data series is summarized in Table \label{tab:uptake}.

\begin{table}
    \centering
    \begin{tabular}{cccccc}
\hline
         & $\log (A\, [length^{-\zeta}mass^{-\omega}])$ & $\phi$ & $\zeta$ & $\omega$ & $R^2$ \\
\hline
    $\log(k_p)$ (Brain) & 4.590	& -0.664 & -0.509 & 0.200 & 0.954 \\
    $\log(k_p)$ (Gill)  & 6.089 & -0.452 & -0.699 & 0.273 & 0.864 \\
    $\log(k_p)$ (Liver) & 6.161 & -0.621 & -0.755 & -0.213 & 0.876 \\
    $\log(k_p)$ (Gut)   & 5.904 & -0.527 & -0.548 & 0.376 & 0.953 \\
\hline
         & $\log (A\, [time\times length^{-\zeta}mass^{-\omega}])$ &  & & & \\
\hline
    $\log(t_c)$ (Brain) & 0.476 & 0.806 & 0.289 & 0.251 & 0.822 \\
    $\log(t_c)$ (Gill)  & 1.811 & 1.089 & 0.022 & 0.349 & 0.791 \\
    $\log(t_c)$ (Liver) & 1.964 & 0.853 & -0.046 & -0.172 & 0.791 \\
    $\log(t_c)$ (Gut)   & 1.953 & 0.953 & 0.116 & 0.603 & 0.761 \\
    \end{tabular}
    \caption{Power law fits of the enrichment factor $k_p$ and the characteristic time $t_c$ of each organ studied, and their corresponding Pearson correlation index.}
    \label{tab:uptake}
\end{table}

\subsection*{Correlations between nanoplastic bioaccumulation, vascular volume, proteins, and lipids}

\begin{figure}
  \centerline{\includegraphics[width=\textwidth]{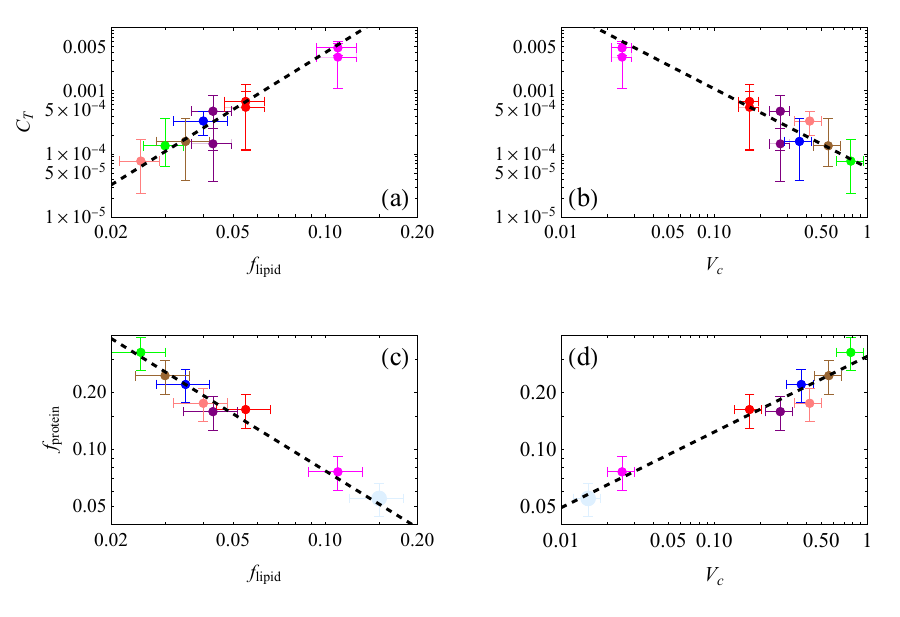}}
  \caption{(a) $C_T$ in seven different organs, as function of the lipid fraction $f^{\text{lipid}}$. Magenta: Brain cortex; red: kidneys; purple: liver. \cite{nihart2025}. Solid colors correspond to 2024, while half opacity colors are for 2016 data. Blue, brown and green: coronary, carotid and aorta arteria, respectively \cite{Liu2024}. Black dashed line is $ C_T = 4 f_{\mathrm{lipid}}^3$. (b) Same ordinate and NP data as in (a), as a function of $V_c$. Black dashed line is $ C_T = 6\times 10^{-5} {V_c}^{-1.2}$. (c) Protein fraction $f_{\mathrm{prot}}$ of the seven organs in (a) and (b) plus brain white matter (light blue), as function of lipid fraction $f_{\mathrm{lipid}}$. Dashed line is $f_{\mathrm{prot}} = 0.0077 f_{\mathrm{lipid}}^{-1}$. (d) Same ordinate as (c), as function of the vascular volume $V_c$. Dashed line is $f_{\mathrm{prot}} = 0.31 {V_c}^{0.4}$.}
\label{fig:SI1}
\end{figure}

Figure~\ref{fig:SI1} shows the concentration of MNPs in selected human tissues, determined by pyrolysis–GC/MS across multiple organs analyzed post mortem \cite{nihart2025,Hu2024,Liu2024}. The panels report pairwise correlations between the measured variables: MNP mass fraction, lipid fraction ($C_{\mathrm{lipids}}$), capillary volume fraction ($V_c$), and protein fraction ($C_{\mathrm{prot}}$). A striking trend emerges across organs, with MNP concentration scaling approximately with the cube of lipid fraction, $C_{\mathrm{lipids}}^{3}$ (panel~a), and exhibiting an inverse dependence on capillary volume fraction of the form $V_c^{-1.2}$ (panel~b). The well-established inverse relationship between lipid and protein fractions is also recovered (panel~c) and is consistent with the strong correlation between protein fraction and capillary volume fraction (panel~d), reflecting the predominance of collagen-rich vascular structures. Although this set of organs is not exhaustive, all reported correlations are statistically significant, with associated $p$-values below $10^{-3}$.

Beyond pairwise correlations, Fig.~\ref{fig:SI1} supports a mechanistic interpretation of organ-specific nanoplastic accumulation. Lipid fraction emerges as the dominant control variable because lipid-rich tissues act as effective capacitive reservoirs for hydrophobic polymers, enhancing both partitioning and residence time. The observed cubic scaling suggests that lipid content controls not only the equilibrium storage capacity but also the effective retention time through repeated systemic recirculation prior to elimination. In contrast, the inverse dependence on capillary volume fraction reflects clearance efficiency: highly vascularized, protein-rich organs are more strongly coupled to systemic elimination pathways and therefore exhibit reduced chronic retention. In liver, fenestrated capillaries enhance clearance efficiency.

The strong correlations among vascular volume fraction, protein fraction, and lipid fraction indicate pronounced multicollinearity among structural organ traits, encoding overlapping aspects of organ bio-architecture. Within this interdependent set of variables, however, only lipid fraction retains a direct and mechanistically explainable correlation with chronic MNP concentration, consistent with the hydrophobic nature of polystyrene. This hierarchy justifies the reduced description adopted here, in which tissues behave as capacitive reservoirs whose effective storage is governed by lipid content, coupled to the systemic compartment through a resistive element associated with the endothelial glycocalyx–endothelium complex. 

\subsection*{Theoretical support to the lipid cubic law}

The dominance of the lipid fraction to justify the cubic law is theoretically supported in the following.

\subsubsection*{Geometric Formulation and Mass Balance}
Consider a Representative Volume Element (RVE) of parenchyma with volume $V_{RVE} \sim R_1^3$, containing a single spherical nanoplastic particle of radius $R_0$. The tissue volume concentration $C_T$ (mass fraction) is defined by the scaling:
\begin{equation}
    C_T \propto \left( \frac{R_0}{R_1} \right)^3
    \label{eq:CT_scaling}
\end{equation}
Within this RVE, the total volume of available lipids is given by $V_{\text{lipids}} = f_{\text{lipid}} R_1^3$. We assume that the NP surface recruits a lipid corona of thickness $\delta$, resulting in a corona volume $V_c \approx 4\pi R_0^2 \delta$. The recruitment of these lipids is governed by a collecting efficiency $\epsilon$, which relates the volume of the corona to the available lipid pool:
\begin{equation}
    R_0^2 \delta \sim \epsilon \cdot f_{\text{lipid}} \cdot R_1^3
    \label{eq:mass_balance}
\end{equation}

\subsubsection*{Orthogonal Effect I: Affinity-Driven Efficiency}
In the physics of polymer-solvent interactions, the compatibility between the polymer and its surrounding medium is typically described by the Flory-Huggins interaction parameter $\chi$. While $\chi$ is often treated as a constant in dilute solutions, in dense biological matrices, the thermodynamic drive for interfacial wetting depends on the local concentration of the solvating species \cite{flory1953principles, russell2015characterization}.

As the lipid fraction $f_{\text{lipid}}$ increases, the chemical potential of the parenchyma becomes increasingly lipophilic, lowering the energy barrier for lipid attachment to the hydrophobic NP surface. Consequently, the collecting efficiency $\epsilon$—representing the probability of a successful recruitment event per unit area—is not a constant but is proportional to the density of the available lipid solute:
\begin{equation}
    \epsilon \propto f_{\text{lipid}}
    \label{eq:epsilon_scaling}
\end{equation}

\subsubsection*{Orthogonal Effect II: Geometric Confinement}
The maximum growth of the lipid corona is physically limited by the proximity of neighboring particles. To model this, we adopt the cell models developed by Happel and Kuwabara for multi-particle systems \cite{happel1958viscous, kuwabara1959forces}. These models define an ``influence zone'' around a particle whose radius $R_1$ scales with the particle volume fraction $\phi$ as $R_1 \propto R_0 \phi^{-1/3}$.

In a chronically bioaccumulated tissue, we treat the RVE as a \emph{Wigner-Seitz cell} \cite{wigner1933constitution}, where the corona thickness $\delta$ is limited by the distance to the cell boundary. This geometric confinement implies that $\delta$ must scale with the inter-particle distance $L \sim R_1$:
\begin{equation}
    \frac{\delta}{R_0} \sim \frac{R_1}{R_0} \propto C_T^{-1/3}
    \label{eq:delta_scaling}
\end{equation}
This represents a ``saturation layer'' where the spatial influence of an NP is constrained by the packing density of the accumulated polymer.

\subsubsection*{Closure of the Cubic Scaling Law}
By substituting the affinity scaling (Eq. \ref{eq:epsilon_scaling}) and the geometric confinement scaling (Eq. \ref{eq:delta_scaling}) into the mass balance (Eq. \ref{eq:mass_balance}), we obtain:
\begin{equation}
    R_0^2 \left( R_0 C_T^{-1/3} \right) \sim (f_{\text{lipid}} \cdot f_{\text{lipid}}) R_1^3
\end{equation}
Using the fundamental relation from Eq. \ref{eq:CT_scaling} ($R_1^3 \propto R_0^3 / C_T$), the expression becomes:
\begin{equation}
    R_0^3 C_T^{-1/3} \sim f_{\text{lipid}}^2 \left( \frac{R_0^3}{C_T} \right) \Longrightarrow C_T \propto f_{\text{lipid}}^3
\end{equation}
%
This derivation strongly suggests that the super-linear bioaccumulation of nanoplastics in the human brain is the emergent result of synergistic quadratic affinity and linear geometric packing effects.

\subsection*{Incorporating a size spectrum: superposition over $d$ with $p(d)$}
\label{si:size_spectrum}

The S2CT model is linear in the driving concentration and therefore admits a \emph{superposition principle} over independent particle-size classes. Let $d$ denote an effective particle diameter (or an equivalent size parameter for non-spherical particles, see below), and let $p(d)$ be the environmental probability density function (PDF) of particle sizes in the relevant exposure medium (water/food/air), normalized on $[d_{\min},d_{\max}]$ as
\begin{equation}
\int_{d_{\min}}^{d_{\max}} p(d)\,\mathrm{d}d = 1.
\end{equation}
If the environment provides a \emph{size-resolved} ambient concentration $C_w(d)$ (mass fraction, wet-weight basis, or an equivalent consistent measure), then the total tissue burden is obtained by integrating the contribution of each size class:
\begin{equation}
C_T(t) \;=\; \int_{d_{\min}}^{d_{\max}} C_T(t;d)\,\mathrm{d}d,
\qquad
C_S(t) \;=\; \int_{d_{\min}}^{d_{\max}} C_S(t;d)\,\mathrm{d}d.
\label{eq:superposition_CT_CS}
\end{equation}
When only a total ambient concentration $C_w$ is available, one may write a partition $C_w(d)=C_w\,p(d)$ (i.e., the ambient concentration is distributed across sizes according to $p$), or, more generally, $C_w(d)=C_w\,f(d)$ where $f$ is a normalized \emph{mass-weighted} (or number-weighted) spectrum, depending on how $C_w$ is defined experimentally. In what follows we keep $C_w(d)$ explicit.

\subsubsection*{Markovian (effective) uptake with $k_p(d)$ and $t_c(d)$}
In the effective Markovian uptake regime used in the main text, each size class follows the same normalized kernel but with size-dependent effective parameters:
\begin{equation}
C_T(t;d) \;=\; C_{T0}(d)\; y_{Tu}\!\left(\frac{t}{t_c(d)};\eta(d)\right),
\qquad
C_{T0}(d) \equiv k_p(d)\,C_w(d),
\label{eq:CTd_markov_uptake}
\end{equation}
where $y_{Tu}$ is the normalized uptake kernel (e.g., Eq.~(12) in the main text, or its $\eta\!\to\!0$ form), $t_c(d)=\lambda_{Te}(d)^{-1}$ is the characteristic tissue time, and $\eta(d)=\lambda_{Se}(d)/\lambda_{Te}(d)$ is the systemic excretion capacity (often effectively small in the analyzed datasets).

If the scale-free fits provide power laws (as in Table~1 of the main text),
\begin{equation}
k_p(d;C_w,W)=A_k\,C_w^{\phi_k}\,d^{\xi_k}\,W^{\omega_k},\qquad
t_c(d;C_w,W)=A_t\,C_w^{\phi_t}\,d^{\xi_t}\,W^{\omega_t},
\label{eq:kp_tc_powerlaws}
\end{equation}
then inserting Eq.~\eqref{eq:CTd_markov_uptake} into Eq.~\eqref{eq:superposition_CT_CS} yields the size-spectrum tissue concentration as the explicit $d$-integral
\begin{equation}
C_T(t)
=\int_{d_{\min}}^{d_{\max}}
\Big[A_k\,C_w(d)^{\,1+\phi_k}\,d^{\xi_k}\,W^{\omega_k}\Big]\;
y_{Tu}\!\left(\frac{t}{A_t\,C_w(d)^{\phi_t}\,d^{\xi_t}\,W^{\omega_t}};\eta(d)\right)\,\mathrm{d}d.
\label{eq:CT_spectrum_markov}
\end{equation}
Equation~\eqref{eq:CT_spectrum_markov} is the sought ``convolution across the whole $d$ spectrum'': the effective ``phantom'' initial concentration $C_{T0}(d)=k_p(d)\,C_w(d)$ acts as the amplitude of each size class, while the kinetics are stretched by the size-dependent timescale $t_c(d)$. In the common low-capacity regime $\eta(d)\!\to\!0$, one may replace $y_T^{(u)}$ by its asymptotic closed form to obtain a fully explicit integrand. For dimensional consistency, the prefactor $A_k$ in Eq.~(\ref{eq:CT_spectrum_markov}) must have units $[\text{length}^{\phi_k-\xi_k}\text{mass}^{-\omega_k}]$ to ensure that $C_T(t)$ retains the correct dimensionality of concentration (mass fraction).

Eq.~(\ref{eq:CT_spectrum_markov}) can be simplified for the long-term asymptotic regime where tissue concentration approaches steady state. In this limit, the normalized uptake kernel approaches its equilibrium value $y_{Tu} \to 1/\eta$, and the size-integrated tissue concentration reduces to:
\begin{equation}
    C_T(t)=\frac{A_k \langle d\rangle^{\xi_k-\phi_k}W^{\omega_k}C_w^{\phi_k + 1}}{\eta} \int_{d_{\text{min}}}^{d_{\text{max}}} f_y^{\phi_k +1}(y) y^{\xi_k}\text{d} y,
\label{eq:CT_asymptotic_spectrum}
\end{equation}
where $f_y(y)=\langle d \rangle f(d)$ is the normalized probability density function, and $y=d/\langle d \rangle$ is the particle diameter normalized by the mean diameter. For a number-weighted distribution $f(d)$, the mean number diameter is defined as:
\begin{equation}
    \langle d \rangle = \int_{d_{\text{min}}}^{d_{\text{max}}} d~f(d) \text{d}d.
\label{eq:mean_diameter}
\end{equation}

To illustrate the practical application of this framework, we consider a representative environmental particle size distribution. Fragmentation-driven nanoplastic size spectra observed in natural environments are well approximated by broad, skewed distributions with power-law cores and exponential tails \cite{Collins2025}. A flexible parametric form capturing these features is the Generalized Inverse Gaussian (GIG) distribution, whose number-normalized dimensionless form is:
\begin{equation}
f_y(y)=\frac{\beta ~y^{\alpha-1}K_{\frac{\alpha+1}{\beta}}(\gamma)^\alpha}{2~K_{\frac{\alpha}{\beta}}(\gamma)^{\alpha+1}} \exp\left(-\gamma\left(\left( \frac{K_{\frac{\alpha+1}{\beta}}(\gamma)}{K_{\frac{\alpha}{\beta}}(\gamma)}\right)^{\beta}y^{\beta} + \left( \frac{K_{\frac{\alpha+1}{\beta}}(\gamma)}{K_{\frac{\alpha}{\beta}}(\gamma)}\right)^{-\beta}y^{-\beta}\right)/2\right),
\label{eq:GIG_distribution}
\end{equation}
where $K_\nu(\gamma)$ denotes the modified Bessel function of the second kind of order $\nu$, and the parameters $\alpha$, $\beta$, and $\gamma$ control the shape, symmetry, and width of the distribution, respectively.

Despite its apparent mathematical complexity, Eq.~(\ref{eq:GIG_distribution}) exhibits a transparent physical structure: it describes a power-law distribution with exponent $\alpha-1$ centered on the mean number diameter $\langle d \rangle$, modulated by exponential decay at both large and small diameters. This functional form has been shown to reproduce fragmentation-driven size spectra observed in marine and freshwater environments with high fidelity \cite{Collins2025}, with the key advantage of continuous validity over the entire domain $0 < d < \infty$, avoiding the artificial cutoffs often required in simpler parameterizations. To demonstrate the quantitative impact of size polydispersity, we evaluate Eq.~(\ref{eq:CT_asymptotic_spectrum}) using parameter values consistent with environmental observations. Following the fragmentation scaling exponent $\alpha = -1$ repeatedly observed by Collins et al.~\cite{Collins2025}, combined with a moderately rapid exponential decay at distribution tails ($\beta = 2$) and a characteristic width spanning three orders of magnitude ($\gamma = 0.001$), the dimensionless size integral in Eq.~(\ref{eq:CT_asymptotic_spectrum}) evaluates to 3.384 for brain and 3.074 for liver. More generally, across the range of realistic distribution parameters characteristic of environmental nanoplastic populations, this integral consistently yields values of order unity, typically in the range 2–5.

The final predicted long-term tissue concentration is then obtained by direct substitution of the organ-specific power-law exponents $\{\phi_k, \xi_k, \omega_k\}$ from Table 1, together with representative values for mean particle diameter $\langle d \rangle$, body mass $W$, ambient concentration $C_w$, and excretion capacity $\eta$, into Eq.~(\ref{eq:CT_asymptotic_spectrum}). This procedure yields quantitative steady-state burden estimates that explicitly account for the full environmental size distribution rather than relying on monodisperse approximations, thereby providing a more realistic representation of actual exposure scenarios.

\subsubsection*{Fractional (general) formulation: Mittag--Leffler kernels integrated over $d$}
For completeness, the same size-superposition applies to the full fractional sequential model, because the governing equations are linear for each $d$:
\begin{align}
{}^{C}\!D_t^{\alpha_S}\,C_S(t;d) &= \lambda_{Su}(d)\,C_w(d)-\lambda_{Se}(d)\,C_S(t;d),\\
{}^{C}\!D_t^{\alpha_T}\,C_T(t;d) &= \lambda_{Tu}(d)\,C_S(t;d)-\lambda_{Te}(d)\,C_T(t;d).
\end{align}
For simplicity, with $C_S(0;d)=C_T(0;d)=0$, the systemic solution is
\begin{equation}
C_S(t;d)=\lambda_{Su}(d)\,C_w(d)\,t^{\alpha_S}\,E_{\alpha_S,\alpha_S+1}\!\big(-\lambda_{Se}(d)\,t^{\alpha_S}\big),
\end{equation}
and the tissue solution is the Mittag--Leffler relaxation kernel convolved with $C_S$:
\begin{equation}
C_T(t;d)=\lambda_{Tu}(d)\int_0^{t}(t-\tau)^{\alpha_T-1}\,
E_{\alpha_T,\alpha_T}\!\big(-\lambda_{Te}(d)\,(t-\tau)^{\alpha_T}\big)\;C_S(\tau;d)\,\mathrm{d}\tau.
\label{eq:CTd_fractional}
\end{equation}
The size-spectrum tissue concentration then follows by integrating Eq.~\eqref{eq:CTd_fractional} over $d$:
\begin{equation}
C_T(t)=\int_{d_{\min}}^{d_{\max}} C_T(t;d)\,\mathrm{d}d,
\end{equation}
or, equivalently, by exchanging the order of integration to obtain a \emph{double convolution} in time and size:
\begin{equation}
C_T(t)=\int_{0}^{t}\left[\int_{d_{\min}}^{d_{\max}}
\lambda_{Tu}(d)\,(t-\tau)^{\alpha_T-1}\,
E_{\alpha_T,\alpha_T}\!\big(-\lambda_{Te}(d)\,(t-\tau)^{\alpha_T}\big)\;C_S(\tau;d)\,\mathrm{d}d\right]\mathrm{d}\tau.
\label{eq:CT_time_size_doubleconv}
\end{equation}
This representation makes explicit how a broad size distribution ``smears'' the single-$d$ Mittag--Leffler response into an effective kernel that generally no longer retains a pure single-exponent (or single-order) form.

\subsubsection*{Remark on non-spherical particles (optional mapping)}
If the environmental population includes non-spherical particles (e.g., cylinders/fibers), the same construction applies by replacing $d$ with an equivalent size parameter $d_{\mathrm{eq}}$
(e.g., same volume-equivalent diameter, same hydrodynamic diameter, or the minor axis for barrier-limited transport). Then $p(d)$ is replaced by the corresponding PDF of $d_{\mathrm{eq}}$ and all expressions above remain unchanged in form.

Once $k_p$ and $t_c$ (or $\lambda_{Te}$) are allowed to depend on $d$ through the empirically determined scale-free laws, linearity implies that the tissue concentration for a realistic environment with a size spectrum is the $d$-superposition of single-size responses, \eqref{eq:CT_spectrum_markov} (effective Markovian) or \eqref{eq:CT_time_size_doubleconv} (fractional). This provides a direct route to propagate an environmental size PDF $p(d)$ into predicted tissue burdens without introducing new ad hoc parameters beyond the measured/assumed $C_w(d)$ and the already-fitted scaling exponents.

\begin{table}
\centering
\renewcommand{\arraystretch}{1.2}
\caption{Symbols and parameters used in the S2CT model.}
\begin{tabular}{lll}
\hline
\textbf{Symbol} & \textbf{Definition} & \textbf{Units / Notes} \\
\hline
$C_w$ & Environmental micro/nanoplastic (MNP) concentration & ppm (mass fraction) \\
$C_S(t)$ & Systemic or blood-phase concentration & ppm \\
$C_{S0}$ & Initial systemic concentration & ppm \\
$C_S(\infty)$ & Equilibrium systemic concentration ($C_S(\infty)\equiv C_{S\infty}$) & ppm \\
$C_T(t)$ & Tissue/parenchymal concentration & ppm \\
$C_{T0}$ & Initial tissue concentration & ppm \\
$C_T(\infty)$ & Equilibrium tissue (organ) concentration ($C_T(\infty)\equiv C_\infty$) & ppm \\
$d$ & nanoparticle diameter & nm \\
$f_{\mathrm{lipid}}$ & Lipid mass fraction in tissue (wet weight) & dimensionless \\
$K_{S}$ & Systemic partition coefficient or enrichment factor ($C_{S\infty}/C_w$) & dimensionless \\
$K_{T}$ & Tissue partition coefficient or enrichment factor ($C_{T\infty}/C_{S\infty}$) & dimensionless \\
$k_d$ & Tissue concentration factor at depuration, $\frac{\lambda_{Tu}}{\lambda_{Te}}\frac{C_S(t_p)}{C_T(t_p)}$ & dimensionless \\
$k_p$ & Generic organ enrichment factor ($C_{T0}/C_w$) & dimensionless \\
$P_u,\,P_e$ & Generalized fractional transport coeffs. (uptake, elimination) & time$^{-1}$ \\
$R_0$ & Capillary radius & length \\
$R_1$ & Mean parenchymal domain radius per capillary & length \\
$t_p$ & Peak concentration time (since initiation of exposure) & time \\
$V_c$ & Vascular volume fraction ($R_0^2 /R_1^2$) & dimensionless \\
$W$ & Average body mass of a vertebrate species & gr \\
$y_{(S,T)u}$ & Normalized (tissue, systemic) uptake concentration & dimensionless \\
$y_{(S,T)d}$ & Normalized (tissue, systemic) depuration concentration & dimensionless \\
$\lambda_{Su},\,\lambda_{Se}$ & Systemic rate coefficients & time$^{-1}$ \\
$\lambda_{Tu},\,\lambda_{Te}$ & Tissue fract. rate coefficients; $\lambda_{Tu,Te}=P_{u,e}R_0/(R_1^2-R_0^2)$ & time$^{-1}$ \\
$\eta$ & Excretion capacity $\lambda_{Se}/\lambda_{Te}$ & dimensionless \\
$\xi_d$ & Depuration coefficient $k_d(1-\eta)^{-1}$ & dimensionless \\
\hline
\end{tabular}
\label{tab:symbols}
\end{table}

\end{document}